\documentclass[final,5p,twocolumn]{elsarticle}
\usepackage{graphicx}
\usepackage{vmargin} 
\setmarginsrb{1.5 cm}{1 cm}{1.5 cm}{1 cm}{1 cm}{1 cm}{1 cm}{1 cm}
\usepackage{fancyhdr}
\usepackage{parskip} 
\usepackage{tikz}
\usetikzlibrary{shapes.geometric} 
\usetikzlibrary{fit} 
\usepackage[edges]{forest} 
\definecolor{folderbg}{RGB}{124,166,198}
\definecolor{folderborder}{RGB}{110,144,169}
\def\Size{4pt}
\tikzset{
    folder/.pic={
        \filldraw [draw=folderborder, top color=folderbg!50, bottom color=folderbg] (-1.05*\Size,0.2\Size+5pt) rectangle ++(.75*\Size,-0.2\Size-5pt);
        \filldraw [draw=folderborder, top color=folderbg!50, bottom color=folderbg] (-1.15*\Size,-\Size) rectangle (1.15*\Size,\Size);
    },
    file/.pic={
        \filldraw [draw=folderborder, top color=folderbg!5, bottom color=folderbg!10] (-\Size,.4*\Size+5pt) coordinate (a) |- (\Size,-1.2*\Size) coordinate (b) -- ++(0,1.6*\Size) coordinate (c) -- ++(-5pt,5pt) coordinate (d) -- cycle (d) |- (c);
    },
    database/.style={
        cylinder,aspect=0.5,draw,rotate=90, path picture={
            \draw (path picture bounding box.160) to[out=180,in=180] (path picture bounding box.20);
            \draw (path picture bounding box.200) to[out=180,in=180] (path picture bounding box.340);
        }
    }
}
\forestset{
    declare autowrapped toks={pic me}{},
    pic dir tree/.style={
        for tree={
            folder,
            font=\small\ttfamily,
            grow'=0,
        },
        before typesetting nodes={for tree={edge label+/.option={pic me}},}
    },
    pic me set/.code n args=2{
        \forestset{
            #1/.style={
                inner xsep=1\Size,
                pic me={pic {#2}}
            }
        }
    },
    pic me set={directory}{folder},
    pic me set={file}{file},
}
\usepackage{listings}
\definecolor{codegreen}{rgb}{0,0.6,0}
\definecolor{backcolour}{rgb}{0.95,0.95,0.92}
\lstdefinestyle{mystyle}{
    backgroundcolor   = \color{backcolour},
    commentstyle      = \color{codegreen},
    keywordstyle      = \color{purple},
    numberstyle       = \tiny\color{gray},
    stringstyle       = \color{blue},
    basicstyle        = \ttfamily\scriptsize,
    breakatwhitespace = true,
    breaklines        = true,
    keepspaces        = false,
    numbers           = left,
    numbersep         = 5pt,
    showspaces        = false,
    showstringspaces  = false,
    showtabs          = false,
    tabsize           = 2,
    postbreak         = \raisebox{0ex}[0ex][0ex]{\ensuremath{\hookrightarrow}}
}
\lstdefinelanguage{custom}{
   morekeywords={gradle,git,localhost:8080,JAVA_HOME,sh,cd,build,x,path,PATH,sudo,vi,DERBY_HOME,tar,wget,javac,java,J2REDIR,J2SDKDIR,MEDIA_DRIVER_DIR,SOFTWARE_PATH,DATA_PATH,[path]},
   sensitive=true,
   morecomment = [l]\#,
}[keywords, comments]
\lstdefinelanguage{Julia}{
    morekeywords = {abstract, break, case, catch, const, continue, do, else, elseif, end, export, false, for, function, immutable, import, importall, if, in, macro, module, otherwise, quote, return, switch, true, try, type, typealias, using, while},
    sensitive = true,
    alsoother = {\$},
    morecomment = [l]\#,
    morecomment = [n]{\#=}{=\#},
    morestring = [s]{"}{"},
    morestring = [m]{'}{'},
}[keywords, comments, strings]
\definecolor{tablegreen}{HTML}{06B400}
\definecolor{tablered}{HTML}{B40000}
\definecolor{tableyellow}{HTML}{E4F11D}
\definecolor{tableblue}{HTML}{000BB4}
\definecolor{tablepurple}{HTML}{980080}
\usepackage{booktabs} 
\usepackage[colorlinks = true, allcolors = blue]{hyperref} 
\usepackage{float} 
\usepackage{comment} 
\usepackage[caption=true]{subfig} 
\usepackage{amsmath}
\usepackage{amssymb} 
\usepackage{fancyvrb} 
\usepackage{fontawesome} 
\usepackage[shortlabels]{enumitem} 
\usepackage{colortbl} 
\usepackage{pifont} 
\usepackage{adjustbox} 
\usepackage{bm} 
\usepackage[linesnumbered,ruled,vlined]{algorithm2e} 
\usepackage{cleveref} 
\numberwithin{equation}{section}
\def\mystrut(#1,#2){\vrule height #1 depth #2 width 0pt}

\newcolumntype{C}[1]{%
   >{\mystrut(3ex,2ex)\centering}%
   p{#1}%
   <{}}

\newlength{\imagewidth}
\journal{SoftwareX}

\begin{document}
\begin{frontmatter}
    \title{CoinTossX: An open-source low-latency high-throughput matching engine}
    
    \author[a1]{Ivan Jericevich}
    \ead{jrciva001@myuct.ac.za}
    \author[a2]{Dharmesh Sing}
    \ead{dharmeshsing@gmail.com}
    \author[a1,a2]{Tim Gebbie}
    \ead{tim.gebbie@uct.ac.za}
    
    \address[a1]{Department of Statistical Sciences, University of Cape Town, Rondebosch 7700, South Africa}
    \address[a2]{Department of Computer Science and Applied Mathematics, University of the Witwatersrand, Johannesburg 2000, South Africa}
    
    \begin{abstract}
        We deploy and demonstrate the CoinTossX low-latency, high-throughput, open-source matching engine with orders sent using the Julia and Python languages. We show how this can be deployed for small-scale local desk-top testing and discuss a larger scale, but local hosting, with multiple traded instruments managed concurrently and managed by multiple clients. We then demonstrate a cloud based deployment using Microsoft Azure, with large-scale industrial and simulation research use cases in mind. The system is exposed and interacted with via sockets using UDP SBE message protocols and can be monitored using a simple web browser interface using HTTP. We give examples showing how orders can be be sent to the system and market data feeds monitored using the Julia and Python languages. The system is developed in Java with orders submitted as binary encodings (SBE) via UDP protocols using the Aeron Media Driver as the low-latency, high throughput message transport. The system separates the order-generation and simulation environments {\it e.g.} agent-based model simulation, from the matching of orders, data-feeds and various modularised components of the order-book system. This ensures a more natural and realistic asynchronicity between events generating orders, and the events associated with order-book dynamics and market data-feeds. We promote the use of Julia as the preferred order submission and simulation environment.
    \end{abstract}
\end{frontmatter}

\tableofcontents

\section{Introduction \label{sec:Introduction}}
A complete study of the market microstructure of the Johannesburg Stock Exchange (\href{https://www.jse.co.za/}{JSE}) is not possible without access to their matching engine\footnote{A matching engine is component of an exchange that matches buy and sell orders according to the rules of the exchange.}. Studying market microstructure is challenging due to the various changes in the market, regulation and technology. However, most of the current literature focuses on analyzing existing exchanges and the building of agent based models. The importance of order matching engines in the trading infrastructure makes these systems of interest not only to computer scientists but also to computational finance and risk management, while the non-linear impact of event-driven processes relating to order matching may provided an impenetrable calibration boundary for agent-based models attempting to empirically relate temporal dynamics with specific agent behaviours \cite{chang2020epps, platt2018can, goosen2020calibrating}.

A trade matching engine is the core software and hardware component of an electronic exchange. It matches up bids and offers to complete trades. Electronic order matching was introduced in the early 1980s in the United States to supplement open outcry trading\footnote{A method of communication between professionals on a stock exchange or futures exchange typically on a trading floor}. Before this, stocks where traded on exchange floors and transaction costs where high. Failures in these systems increased as the frequency and volume on the electronic networks increased. 

Modern matching engines are fully automated and use one or several algorithms to allocate trades among competing bids and offers at the same price. They typically support different order types and have unique APIs or use standard ones\footnote{Vendors include \href{https://www.connamara.com/}{Connamara Systems}, \href{http://www.cinnober.com/}{Cinnober} (acquired by Nasdaq), \href{https://www.aquis.eu/aquis-technologies}{\textbf{Aquis Technologies}} \textbf{(A2X)}, \href{https://www.sia.eu/en}{SIA S.p.A.}, \href{https://www.nasdaq.com/}{Nasdaq}, \href{https://www.match-trade.com/}{Match-Trade}, \href{https://www.millenniumitesp.com}{\textbf{MillenniumIT}} \textbf{(JSE)}, GATElab Ltd (acquired by London Stock Exchange), \href{https://www.eurex.com/ex-en/}{Eurex}, \href{https://www.list-group.com/exchange-technology/}{LIST}, \href{https://www.stellartradingsystems.com/}{Stellar Trading Systems}, \href{https://home.quodd.com/}{Quodd}, \href{https://www.baymarkets.com/}{Baymarkets}, \href{http://marketgridsystems.com/}{Market Grid}, \href{https://www.arqatech.com/en/}{ARQA Technologies}, \href{http://www.kappsoft.com/}{Kappsoft}, \href{https://www.thesystech.com/}{Thesys Technologies}}. As it pertains to trading, latency\footnote{Latency refers to the ability of a system to handle data messages with minimal delay.} directly influences the amount of time it takes for a trader to interact with the market.

Traditionally, to achieve low latency, high-frequency trading has required powerful server hardware appropriately networked in a data center, scaled to accommodate worst-case network traffic scenarios on the busiest trading days. These trading systems must be resilient in the face of network or power failures, requiring expensive redundant hardware as well as offsite data retention \cite{addison2019low}. For example, firms would use co-location, fibre-optic network cables, optimized hardware architectures and other technology to get as close to zero latency as possible \cite{sing2017cointossx}. On the other hand, cloud-based software solutions benefit by being resilient and offering cost-effective, easy scaling --- an advantage that is not offered by traditional trading systems. However, the biggest challenge for latency in a cloud-based environment, and one of the greatest barriers to building high-frequency trading systems in this environment, is the fact that hardware is not co-located within a data center \cite{addison2019low}. That said, a combination of the low latency of traditional matching engines and the resiliency, scalability and availability of cloud-based environments is something that is yet to come about\footnote{There is however a strong argument to place the Order Management System (OMS), that decides what to trade - the selection of parent orders and execution strategies, in the Cloud; while retraining the Execution Management System (EMS), that implements child orders, in close proximity to the matching engine. Here we are moving the matching engine into the Cloud.}.

A low latency high throughput\footnote{High throughput refers to the ability of a system to process a very high volume of data messages.} matching engine does not exist for academics to further their understanding in this field \cite{sing2017jse}. CoinTossX provides an environment for the application of agent-based modelling experiments which would otherwise be prohibitive to undertake in real financial markets due to cost, complexity and other factors. It was for this reason that CoinTossX was developed and why it's applications are studied further here. Here we hope to provide an easy-to-use, openly available, realistic, real-time, simulated trading system that is straight-forward to set-up on multiple platforms. Hence, this paper benefits institutions, academics and others seeking to take advantage of an open-source, resilient, scalable matching engine simulator that may avoid a variety of conflicts of interests related research or insights derived from commercial alternatives.

This paper is structured as follows: \Cref{sec:Literature review} and \Cref{ref:concurrency} outline the existing literature on the topic. \Cref{sec:CoinTossX} is dedicated to an outline of CoinTossX. More specifically, \Cref{sec:Architecture} gives a detailed description of the structure and software construction. \Cref{sec:Functionality} presents the capabilities and features of CoinTossX. \Cref{sec:Testing framework} gives the details of the extensive tests conducted to ensure the system meets the stringent latency and throughput requirements. Supplementary to the testing framework, \Cref{sec:Simulation results} shows the results of the simulation of a simple Hawkes process for generating large volumes of market and limit orders. \Cref{sec:Conclusion} ends with some concluding remarks and, lastly, \ref{sec:Deployment} provides instruction for users to deploy and use the application on their local machine or on a remote server.

\section{Market simulation for testing \label{sec:Literature review}}

Simulating the entire financial market ecosystem for trade strategy testing and risk management is appealing because of the mechanistic complexity of the market structure and costs and the nonlinear feed-backs and interactions that multiple interacting agents bring to the market ecosystem. This type of simulation can be both a financially costly, as well as computational expensive exercise; yet it appears tractable, and subsets of the ecosystem are used for system verification {\it e.g.} in vendor provided test market venues. 

The rapid evolution of software and hardware and the increasing need to reduce transaction costs, system failures and transaction errors has led to important changes in market structure and market architecture. These changes have supported the rise of electronic, algorithmic, and high frequency trading. The specifics are unique to each and every market and regulatory environment. For example, in the South African context the JSE uses the MillenniumIT trading systems following the approach of the London Stock Exchange with the BDA broker-dealer clearing system \cite{jse2020bda} with rules and regulations set by the Financial Markets Act (2012), the JSE rules and directives, and the Financial Intelligence Centre (FIC) Act (2001). However, CoinTossX is fully configurable. Although implemented and tested here using the published and publicly available JSE market rules and test-cases, it can be configured for a rich variety of market structures. 

Many researchers developing trading strategies do not have access to vendor provided testing environments, while those that do often do not want to expose their strategies to competitors in shared testing environments. However, sufficiently realistic multi-agents simulation environments remain illusive, particularly for low liquidity, and collective behaviour based risk-event scenarios \cite{jericevich2020comparing} because a key component remains the realism of the underlying trading agents and their interactions within test markets.

Most agent-based simulation environments and models are over simplified to the extent of not being relevant for realistic science and risk management. Some examples include using a global calendar time to sequentially order trading events, or providing intentional (and sometimes unintentional) market clearing events that synchronise price discovery and information flow with agent interactions in terms of a unique global time -- the loss of high-concurrency and cohersion in favour of tight-coupling for computational convenience. 

There are many examples, in the context of South African markets. \citet{nair2015agent} prototyped a simple matching engine using a single stock and adopting the standard order types and mechanisms associated with a continuous double auction market as specified in the JSE. Although works like this seem to provide a simplified but realistic framework demonstrating the over-all principles of coupling a matching engine with agent-based modeling of a stock market; such frameworks are unlikely to recover realistic dynamics or lead to useful insights when interrogated at the level of asynchronous but high concurrency order-book events and dynamics. High-concurrency and a careful management of the concept of time \cite{chang2020epps} seem a prudent requirement related to both the stylised facts of the market, as argued for in many South African market examples \cite{jericevich2020comparing}, but more importantly, for realistic strategy testing and risk management. 

Here like-for-like message delays, order-book rules, and both asynchronous order matching as well as reactive, asynchronous, and high-concurrency agent (for the rules and behaviours) and actor (for the computational representations and causation models) generation can affect model outcomes and estimation. It is the asynchrony and order-splitting at the agent level that can dominate the emergence of auto-correlations and cross-correlations in various traded assets because there is no single calendar time related mechanisms that generates equilibrium prices, can synchronise information and order-flow, or can co-ordinate machine time events, with sequential calendar time \cite{chang2020epps}.

\section{Realistically simulating high-concurrency} \label{ref:concurrency}

The LMAX Exchange (London Multi Asset Exchange) \cite{thompson2011disruptor} is an FX exchange with an ultra low latency matching engine. \citet{thompson2011disruptor} developed the Disruptor ring buffer for inter-thread concurrency as an alternative to storing events in queues or linked lists. This was in response to the problem that linked-lists could grow and increase the garbage in the system --- causing significant costs to latency and jitter\footnote{Jitter is the deviation from true periodicity of a presumably periodic signal, often in relation to a reference clock signal. This variation in the time between data packets arriving is caused by network congestion.}. At the heart of the disruptor mechanism sits a pre-allocated bounded data structure in the form of a ring-buffer. The Disruptor preallocates memory in a cache-friendly manner which is shared with the consumer. The system was also developed on the JVM and can process up to 25 million messages per second on a single thread with latencies lower than 50 nanoseconds \cite{thompson2011disruptor}.

Recently, \citet{addison2019low} implement a simple foreign exchange (FX) trading system and deploy it to cloud environments from multiple cloud providers (Amazon Web Service, Microsoft Azure and Oracle Cloud Infrastructure), recording network latency and overall system latency in order to assess the capability of public cloud infrastructure in performing low-latency trade execution under various configurations and scenarios. They conclude that sufficiently low latency and controlled jitter can be achieved in a public cloud environment to support security trading in the public cloud \cite{addison2019low}. More specifically, they demonstrate the ability to achieve sub-500 microsecond roundtrip latency --- therefore concluding that it is currently feasible to build a production low-latency, high-frequency trading system in the cloud.

For research in finance, economics and computer science, the importance of having a realistic, flexible, high performance matching engine deployable to different environments can be found in a wide range of fields. In particular, given the whole new range of realism that such a software provides, agent-based computational finance and financial models in general are some of those that may show the greatest potential in terms of the insights to complex systems that can be gained from their application. This is not to mention the additional class/layer of causation that is introduced in the modelling framework by having the complex set rules of the environment/system/architecture be separate/independent from the modelling and decision making processes (at the agent level). In this way more emphasis can be placed on top-down actions and states --- a potential step towards hierarchical causality \cite{wilcox2014hierarchical}.

On this note, CoinTossX is a simulation environment and so the task of producing realistic simulated market dynamics, comparable to those observed in empirical investigations, is left to the user(s). For this purpose the two popular methods usually adopted are mutually exciting Hawkes processes (see \Cref{sec:Testing framework}) and agent-based models (ABMs) \cite{lussange2018bright}.

ABMs provide a bottom-up approach to modelling the actions and interactions of autonomous agents with the aim of assessing their effect on a complex system. Proponents of agent-based models argue that financial markets exhibit many emergent phenomena, and that such phenomena are usually attributed to the interactions and relationships between the agents that make up the system. Successfully calibrating ABMs to financial time series \cite{platt2020calibration} can allow for inference about the factors determining the price behaviour observed in the real world, provided that parameter estimates are sufficiently robust.

In recent years, ABMs became popular as a tool to study macroeconomics — specifically, the impact of trading taxes, market regulatory policies, quantitative easing, and the general role of central banks. ABMs can also play an important role in analysis of the impact of the cross-market structure \cite{lussange2018bright}. Of particular interest is the class of models described by \citet{lussange2018bright} who outline a computational research study of collective economic behavior via agent-based models, where each agent would be endowed with specific cognitive and behavioral biases known to the field of neuroeconomics, and at the same time autonomously implement rational quantitative financial strategies updated by machine learning (and reinforcement learning (RL)). 

ABMs are not without their pitfalls \cite{lebaron2006agent,lussange2018bright,platt2020calibration}: computational cost, validation and calibration challenges, a bias towards inductive decision making, endemic parameter degeneracies, and their tendency to focus on mechanistic behaviours and rules, rather than interaction dynamics and learning . Despite these challenges, their potential to link the micro-level rules of investors behaviour with the macro-behaviour of asset prices in real market is compelling, in part due to the apparent expansive amount of data that is collected from financial markets. However, a key concern remains the scientific costs, or impact, of losing the realism of high-concurrency adaptive interactions between strategic agents in a sufficient reactive framework, for the computational convenience of time-synchronised bottom-up rule based approaches. Here we have decided to first focus on the matching engine framework, and separate it entirely from the agent generation framework; this may ensure that high-concurrency and low latency features of real markets are not lost.  

\section{CoinTossX \label{sec:CoinTossX}}

CoinTossX is a an open-source, high-frequency, low latency, high throughput matching engine for simulating the JSE \cite{sing2017jse, sing2017cointossx, Jericevich2021} (or any market that has well established rules and test cases). The software was developed with Java and open-source libraries and is designed to maximize throughput, minimize latency, and accommodate rapid development of additional functionality. It can be configured for multiple clients, stocks and trading sessions (continuous trading, opening auction, closing auction, intraday auction and volatility auction). The software has been configured to replicate the rules and processes of the JSE. It may allow traders, organizations and academic institutions to test market structure, fragility and dynamics without the cost of live test trading. The software can provide a platform to study price formations in stock exchanges and the interplay between regulators, market structure and dynamics \cite{sing2017jse}. This work also addresses some aspects of the unavailability of data and direct data-feed access from industry, by providing a framework that can be compared to recorded transaction data arising from the actual market system interfaces. 

The system requirements we implemented were obtained from JSE's publicly available technical documentation and test cases \cite{jse2020trading, jse2020native, jse2020market}\footnote{Technical documentation can be found at \href{https://www.jse.co.za/services/technologies/equity-market-trading-and-information-technology-change}{https://www.jse.co.za/services/technologies/equity-market-trading-and-information-technology-change}}.

The eight main components of the simulator are \cite{sing2017jse}:
\begin{enumerate}
  \item \textbf{Stocks} are objects for which clients can send orders and limit order books can be constructed and kept track of.
  \item \textbf{Clients} are computer algorithms that send order events to the simulator. Clients will send order events to the trading gateway (refer to Section \ref{sec:Clients} for more detail).
  \item The \textbf{trading gateway} receives the client request, validates the request and then sends it to the matching component to be processed. It sends updates to the client to indicate the status of the event.
  \item The \textbf{matching engine} processes the events from the Trading Gateway. It manages one or more limit order books. If there is an update to the LOB, it sends updates to the market data gateway.
  \item The \textbf{market data gateway} receives updates from the matching engine component and sends market data updates to all connected clients.
  \item The \textbf{website} receives updates from the market data gateway. It displays the LOBs for each security and allows the user to configure the stocks and clients.
  \item The \textbf{web-event listener} was specifically incorporated for the purpose of reducing pauses to the website due to garbage collection. This single, dedicated writer principle would improve performance of the website as well (refer to the discussion on the website in Section \ref{sec:Architecture}). This component also saves each event to the database
  \item The \textbf{database} stores the LOBs of each stock and is designed to have low memory overhead whilst being efficient in searching, updating and deleting orders (refer to the discussion on off-heap data structures in Section \ref{sec:Architecture}).
\end{enumerate}

\begin{figure}[H]
    \centering
    \resizebox{0.48\textwidth}{!}{
        \begin{tikzpicture}[square/.style = {regular polygon, regular polygon sides = 4}]
            \node [square, draw, inner sep = 0, text width = 1.3cm, align = center] (change trading session tg) at (2,5) {\small Change\\trading\\session};
            \node [square, draw, inner sep = 0, text width = 1.3cm, align = center] (login) at (4,5) {\small Login};
            \node [square, draw, inner sep = 0, text width = 1.3cm, align = center] (log out) at (6,5) {\small Log\\out};
            \node [square, draw, inner sep = 0, text width = 1.3cm, align = center] (order request tg) at (8,5) {\small Order\\request};
            \node[align = center] (tg) at (4.5, 3.5) {\small Trading gateway};
            \node[draw, fit = (tg)(order request tg)(change trading session tg), ultra thick, green] (trading gateway) {};
            \node [ellipse, draw] (client) at (11,1.5) {\small Client};
            \node [ellipse, draw] (user) at (5,-2) {\small User};
            \node[align = center] (web) at (4, 1.5) {\small Website};
            \node [square, draw, inner sep = 0, text width = 1.3cm, align = center] (manage clients) at (2,0) {\small Manage\\clients};
            \node [square, draw, inner sep = 0, text width = 1.3cm, align = center] (manage stocks) at (4,0) {\small Manage\\stocks};
            \node [square, draw, inner sep = 0, text width = 1.3cm, align = center] (view lobs) at (6,0) {\small View\\LOBs};
            \node [square, draw, inner sep = 0, text width = 1.3cm, align = center] (run) at (8,0) {\small Run testing\\framework};
            \node[draw, fit = (manage clients)(web)(run), ultra thick, purple] (website) {};
            \node [square, draw, inner sep = 0, text width = 1.3cm, align = center] (order request me) at (12,5) {\small Order\\request};
            \node [square, draw, inner sep = 0, text width = 1.3cm, align = center] (change trading session me) at (14,5) {\small Change\\trading\\session};
            \node [square, draw, inner sep = 0, text width = 1.3cm, align = center] (market data update me) at (16,5) {\small Market\\data\\update};
            \node[align = center] (me) at (13, 3.5) {\small Matching engine};
            \node[draw, fit = (me)(order request me)(market data update me), ultra thick, blue] (matching engine) {};
            \node [square, draw, inner sep = 0, text width = 1.3cm, align = center] (market data update mdg) at (14,0) {\small Market\\data\\update};
            \node[align = center] (mdg) at (14, 1.5) {\small Market data\\gateway};
            \node[draw, fit = (market data update mdg)(mdg), ultra thick, red] (market data gateway) {};
            \draw[->] (user) -- (manage clients);
            \draw[->] (user) -- (manage stocks);
            \draw[->] (user) -- (view lobs);
            \draw[->] (user) -- (run);
            \draw[->] (client) -- (login);
            \draw[->] (client) -- (log out);
            \draw[->] (client) -- (order request tg);
            \draw[->] (website) -- (change trading session tg);
            \draw[->] (trading gateway) -- (matching engine);
            \draw[->] (market data update me) -- (market data gateway);
            \draw[->] (market data update mdg) -- (website);
            \draw[->] (market data update mdg) -- (client);
        \end{tikzpicture}
    }
    \caption{A High level diagram of the relationship between the components of CoinTossX in terms of end-user functionality \cite{sing2017jse}. \label{fig:cointossx}}
\end{figure}
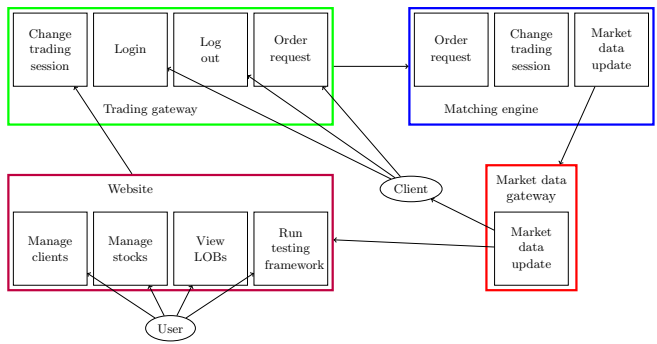

A website is required to monitor and configure the stocks and clients. The relationships between these components are also summarised in Figure \ref{fig:cointossx}. The client sends a login request message to the trading gateway which then validates the client. If the client is already logged in or the username or password is invalid it responds with a reject code message. The client can also send a log out request to the trading gateway. In this case the trading gateway removes the client from the list of connected clients and sends a response to the client to indicate if the log out was successful. The trading gateway logs out clients when it shuts down. When there is a change in the trading session, the website sends a message to the trading gateway which then sends the message to the matching engine. Clients can send an order request or order message to the trading gateway which then validates the message. If the message is valid, it then sends the message to get matched. Valid order types are market, limit, hidden, stop and stop limit. Each order has a time in force (TIF) value, which cannot be amended, that determines how long an order is active until it is executed, deleted or expired (which ever comes first). The engine matches the orders using the matching algorithms based on the trading session. Market data updates are sent to the market data gateway to indicate the changes in the LOB of each stock. Additionally, the website and all clients receive market data updates and internal messages (to monitor the application) from the gateway. A user can create, update and delete clients and stocks. A user can view the limit order book for each stock as a bar chart of all active orders. Tables will show the details of these active orders. A user can start, stop and configure the parameters of the testing framework.

\subsection{Architecture \label{sec:Architecture}}
\begin{figure}[H]
    \centering
    \resizebox{0.48\textwidth}{!}{
        \begin{tikzpicture}
            \node [draw, inner sep = 5, align = center, rounded corners, ultra thick] (clients) at (8,10) {\small Client $\{ 1, \hdots, N \}$ \faGroup};
            \node [draw=orange, inner sep = 5, align = center, rounded corners, ultra thick] (md1) at (2,8) {\small Media driver \faPlug};
            \node [draw=orange, inner sep = 5, align = center, rounded corners, ultra thick] (md2) at (8,8) {\small Media driver \faPlug};
            \node [draw=orange, inner sep = 5, align = center, rounded corners, ultra thick] (md3) at (14,8) {\small Media driver \faPlug};
            \node [draw=green, inner sep = 5, align = center, rounded corners, ultra thick] (tg) at (2,6) {\small Trading \faSignIn\\gateway \faSignOut};
            \node [draw=blue, inner sep = 5, align = center, rounded corners, ultra thick] (me) at (8,6) {\small Matching engine};
            \node[inner sep=0pt, opacity=0.5] (img) at (8,6) {\includegraphics[width=.1\textwidth]{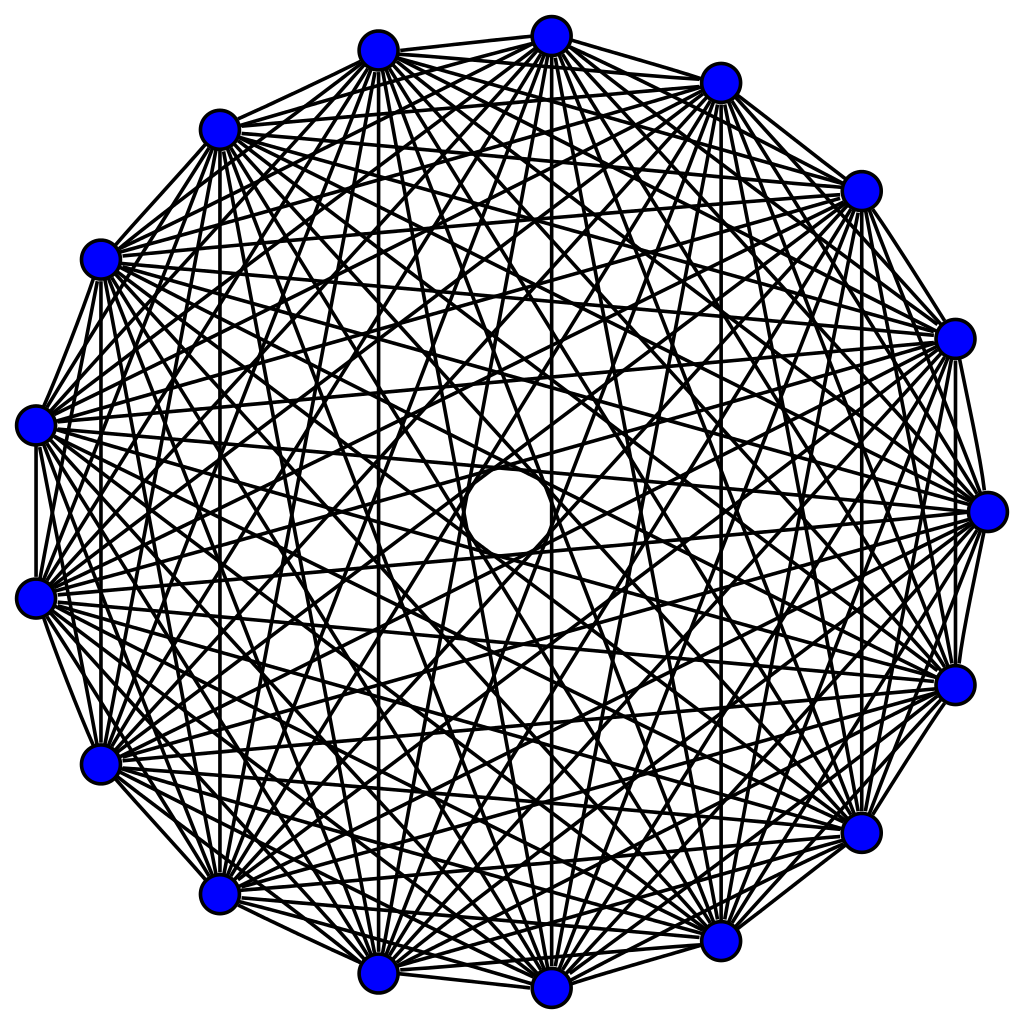}};
            \node [draw=red, inner sep = 5, align = center, rounded corners, ultra thick] (mdg) at (14,6) {\small Market data\\gateway \faLineChart};
            \node [draw=orange, inner sep = 5, align = center, rounded corners, ultra thick] (md4) at (14,4) {\small Media driver \faPlug};
            \node [draw=yellow, inner sep = 5, align = center, rounded corners, ultra thick] (wel) at (10,4) {\small Web event\\listener \faMixcloud};
            \node[database, scale=3] (fs) at (6,4) {}; \node[align = center] at (6, 3.5) {\small File system};
            \node [draw=purple, inner sep = 5, align = center, rounded corners, ultra thick] (web) at (2,4) {\small Website \faInternetExplorer};
            \draw [>=latex,->,thick] (web) -- +(-2,0) |- (md1);
            \draw [>=latex,->,thick] (md3) -- +(2,0) |- (md4);
            \draw [<->,thick] (clients) -| (md3);
            \draw [<->,thick] (clients) -| (md1);
            \draw [<->,thick] (md1) -- (md2);
            \draw [<->,thick] (md2) -- (md3);
            \draw [<->,thick] (tg) -- (md1);
            \draw [<->,thick] (me) -- (md2);
            \draw [<->,thick] (mdg) -- (md3);
            \draw [->,thick] (md4) -- (wel);
            \draw [->,thick] (wel) -- (fs);
            \draw [->,thick] (fs) -- (web);
        \end{tikzpicture}
    }
    \caption{A High level architecture diagram visualising the technical relationship between the software components \cite{sing2017jse}. \label{fig:architecture}}
\end{figure}
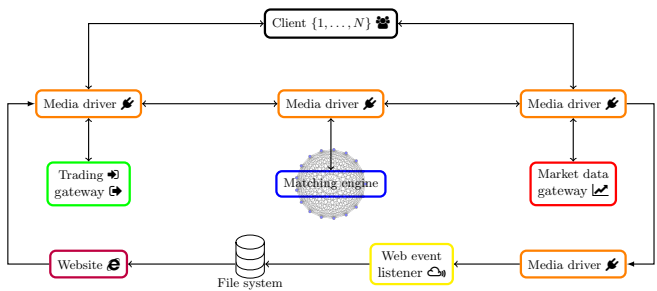

This software is open source and was built to run on different operating systems as well as on a single server or on multiple servers. Implementation in Java means that CoinTossX can be deployed on different hardware configurations (due to the JVM). The goal was to build a matching engine that could achieve low-latencies using industry standard technology. This ensures that the software is portable and can be deployed confidently. The high level architecture of CoinTossX is summarised in Figure \ref{fig:architecture}.

A single market event goes through the following process. A simple binary encoding (SBE) message\footnote{JSE uses text messages between the clients and their matching engine which is inefficient since characters take up more memory and are slower to transmit across the network. From the comparisons made in \cite{sing2017jse}, the SBE message protocol was found to be the fastest way to encode and decode messages; and had the greatest throughput.} will be sent to the trading gateway. The trading gateway will forward the message to the matching engine which will then process it. It will send a message back to the trading gateway to indicate the status of the message. The trading gateway will forward the message back to the client. The matching engine will send an update to the market data gateway if there is a change in the limit order book. The market data gateway will forward the updates to all connected clients and will forward some of the updates to the web event listener. The web event listener saves each event to the file system. The website then reads and displays the data from the file system.

The communication between the website and the file system is done by transmitting data over a network using only User Datagram Protocol (UDP)\footnote{JSE uses TCP for their trading gateway and UDP for their market data gateway.} as opposed to Transmission Control Protocol (TCP). TCP is a heavyweight protocol because it requires three packets to set up a connection and requires a connection to be set up between two applications before data can be transmitted\footnote{Nonetheless, TCP can be reliable because if a message is not received, it will try multiple times to deliver the message. TCP will drop the connection if there are multiple timeouts.}. UDP, on the other hand, was chosen for being lightweight protocol as it does not check the connection or the order of the messages and does not require a connection to be setup before data is transmitted. Data is transmitted irrespective of whether the receiver is ready to receive the message or not. UDP may, however, be unreliable because the sender does not know if the message was delivered. The messages that are received (read as a byte stream) will always be in the order that it is sent. Modularisation is achieved by designing each of the above components independently of one other while allowing communication between them to occur only via exposed ports using these high speed SBE message protocols. This functionality was introduced by assigning each component an IP address and port on which to listen. These ports are specified in the ``properties'' files in the root project directory. Furthermore, each client/user may submit orders from a remote server. So if CoinTossX is deployed to the cloud, one should ensure that the virtual machine allows for these types of inbound port communications.

\begin{figure}[H]
    \centering
    \includegraphics[width=.48\textwidth]{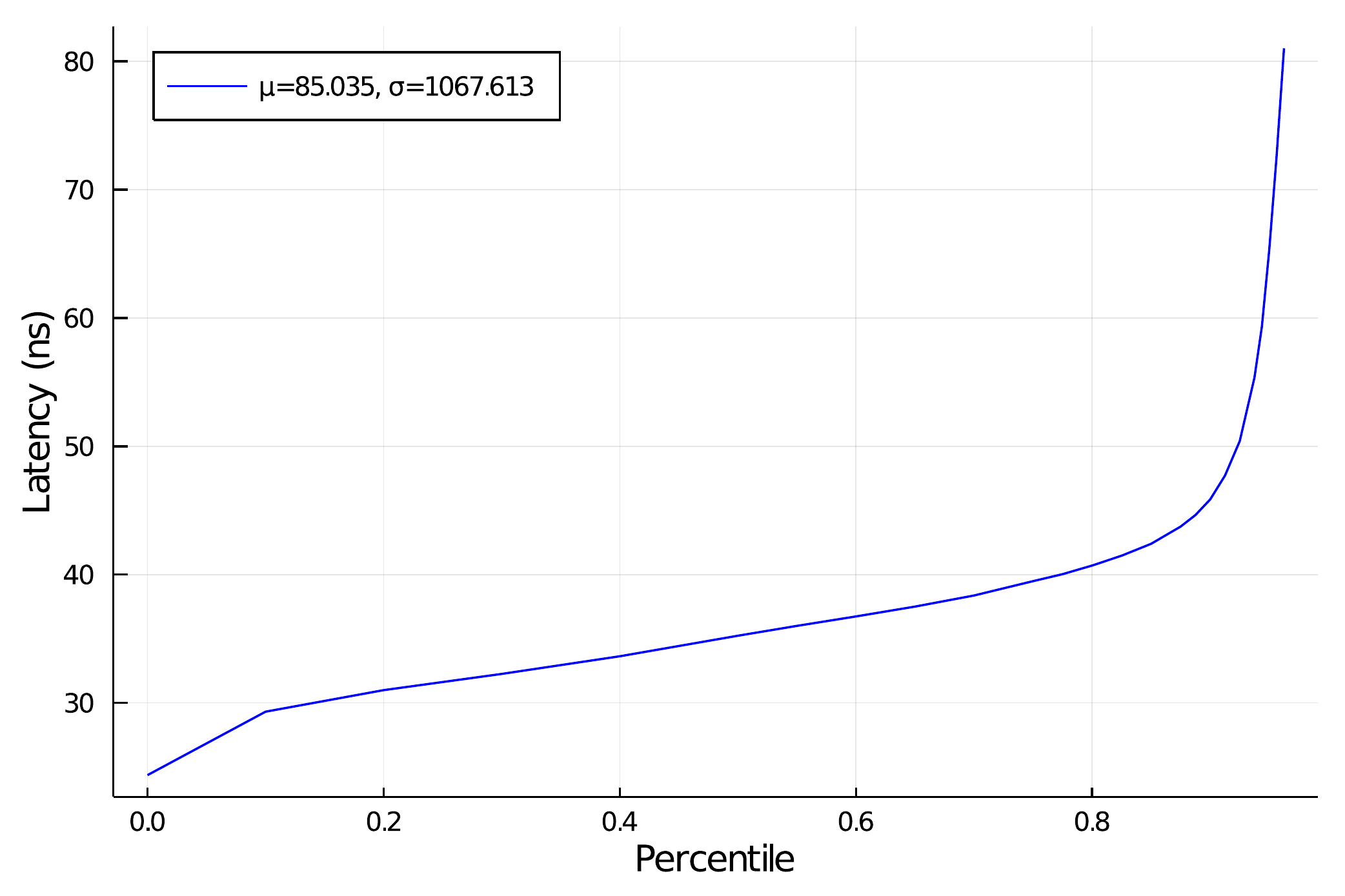}
    \caption{A High Dynamic Range (HDR) Histogram graph showing the latency percentiles of Aeron’s PingPong test. This provides a lower-bound for the achievable latency of CoinTossX as well as for open-source message transport libraries in general \cite{real2021aeron}. \label{fig:aeron}}
\end{figure}

After an extensive comparison of message transport software, the \href{https://github.com/real-logic/aeron}{Aeron media driver} library was chosen for supporting the above protocols being an efficient and reliable UDP unicast, UDP multicast, and IPC message transport for communicating between all the components. Each component has its own media driver to shovel events to and from the component. Aeron is designed to be the highest throughput with the lowest latency possible of any messaging system \cite{real2021aeron} (see Figure \ref{fig:aeron}).

Given this modular design, each component and client may be started and run on seperate, independent servers. As mentioned, components communicate with each other through the media driver via UDP SBE message packets. Therefore. it is essentially possible to have an industrial matching engine by providing each component with its own high performance server.

The matching engine is the component which relies heaviest on the latency and throughput of the system and is therefore designed with these features in mind whilst sticking to the JSE matching engine's architecture. The software was designed such that different matching logic algorithms are in separate Java classes. This allows the logic to be changed to test any variations of the matching logic. The matching engine adopts the price-time priority algorithm during the continuous trading session. That is, for multiple orders occurring at the same price, a market order will be matched with the order having the earliest submission time. The trading session execution times and monitoring was not implemented in the components as this would reduce the throughput and increase the latency.

Limit and market orders submitted during a call auction are not matched immediately. These orders are matched at the end of the call auction at a single price using a price discovery process. The Volume Maximizing Auction Algorithm finds the price that will match the most number of buy and sell orders. More specifically, during both the continuous trading and auction call sessions, the matching engine processes orders as follows. The filter and uncross algorithms run each time the Best Bid or Offer (BBO) changes or every 30 seconds. The filter algorithm uses the heuristic Hill Climber search/optimization algorithm\footnote{The Hill Climber algorithm is an optimization technique that iteratively searches for the solution to a problem by changing one element in each iteration}  to find the optimal volume of hidden limit orders that can be executed. The search will filter out hidden limit orders with MES constraints that are not eligible. After the filtering, specific rules are used to select the orders and price to executed in the crossed region.

As in the JSE, the matching engine uses native message protocols\footnote{Message protocols are the methods by which the exchange communicates with market participants. Native protocols are custom built protocols.} to communicate with clients as opposed to FIX\footnote{FIX is an open standard non-propriety protocol that is used by buy and sell firms, trading platforms and regulators to transmit trade data. The protocol allows organizations to easily communicate domestically and internationally with each other.} (Financial Information eXchange), FAST\footnote{Greater numbers of market participants and therefore volume increased the network latency and lead to market participants not receiving updates in an acceptable time. The FAST protocol reduces latency by encoding and compressing the data before transmission.} (FIX Adapted for STreaming), ITCH\footnote{ITCH is used to publish market data only.} or OUTCH\footnote{OUCH is used to place, amend and cancel orders.}. The matching engine component stores the active orders in memory since storing the active orders on disk would increase the I/O and reduce performance. The data structure for storing the LOB is designed to have a low memory overhead (lists, hash tables, trees, B-tree, B+tree) and be efficient in searching, updating and deleting orders \cite{sing2017jse}. More importantly, the low latency is achieved through the efficient use of CPU cache. The simulator only uses main memory without the use of virtual memory\footnote{Virtual memory is space on the hard disk which is used by an operating system when it requires more memory. The time to fetch data from virtual memory is very slow and is therefore not used.}. The time taken for data to move from the CPU to main memory is reduced by using caches that contain copies of frequently used data from main memory.

The website was developed using \href{https://spring.io/projects/spring-boot}{Spring Boot} and \href{https://wicket.apache.org/}{Apache Wicket} and only has the permission to read from the file system. The original design had the event listener and website in the same Java process. When the website was used or paused because of garbage collection, it affected the receiving and saving of events. Therefore this logic was split into a separate component --- the web event listener. The listener could keep the received events in memory or save it to the file system. Saving the data in memory would be fast, but would require an unknown maximum memory setting. Therefore the data needs to be saved to the file system. Using the MapDB library\footnote{MapDB is an open-source, embedded Java database engine and collection framework. It provides Maps, Sets, Lists, Queues, Bitmaps with range queries, expiration, compression, off-heap storage and streaming.}, an off heap hashmap is used to save the events to memory mapped files. An off heap hashmap stores data outside the Java heap space and is not affected by garbage collection. Off heap data is suited for storing data larger than the current memory and allows sharing of data between JVMs. Memory mapped files allow Java programs to read and write files using only memory while the operating system reads and writes to the file system. This significantly improves performance. The entire file or a part of the file can be loaded into memory. The values in memory will still be written to the file system even if the JVM crashes. The web event listener has read and write permissions. It saves events to the file system but receives events faster than it can save it to file. This issue is solved with the Disruptor by having two threads: one to receive and store events, and the other to save events to the file system \cite{thompson2011disruptor}.

\subsection{Clients \label{sec:Clients}}

Important to the design of this trading system are the clients who send orders to the trading gateway and receive updates from the website and market data gateway. Each client is assigned input and output URLs for both the Native Gateway and Market Data Gateway. These URLs specify the IP addresses (in this case localhost - the user's local machine) to/from which messages will be sent/received as well as the ports on which these components are listening. Therefore each client takes up a number of threads on the machine it runs on --- which can be the same or different from the machine running the other components of the matching engine. So, having clients send orders to the server remotely would free up hardware and improve performance on the matching engine server.

It is also important to note that each client here is a unique trader having its own unique ID, password, ports and login credentials. However, as was done in the testing framework, a single client may still be used to simulate multiple traders. This would be preferable in the case where the client(s) and other components run on the same machine, thereby limiting the computational resources required for simulations. In that case, the client is simply a tool/object for submitting orders to the gateway. In this way one can have a self-contained simulation framework be independent from the actual implementation or sending of orders by, for example, defining an ABM in julia with each agent holding a reference to the client object that sends orders.

The number of client-stock pairs that can be supported in total is only limited to the performance capabilities of the server on which the matching engine is running (see section \ref{sec:Testing framework} for hardware recommendations and performance results with differing number of client-stock pairs). The limits of the number of clients that can be supported by each stock, however, still needs to be explored and measured with respect to different hardware configurations.

\subsection{Functionality \label{sec:Functionality}}
When the user starts CoinTossX, through the website, three main screens can be switched between and displayed. The stock screen shows the stocks configured, the trading session that is active for each stock and a button to view the limit order book of the stock. In the limit order books for each stock a bar chart will be generated to display the active orders. The market data shown on the LOB pages for each stock are snapshots representing the current state of the limit order book. Tables will show the details of the bids, offers, trades and all submitted orders. All data on these pages can be exported. The Hawkes configuration page (see Section \ref{sec:Testing framework}) allows the user to change the values of the Hawkes input data before running the simulation. The simulation page allows the user to stop and start the warmup process and the Hawkes simulation. It shows the status of each client and the active trading session. The clients screen shows the clients that are configured and allows the user to create, update and delete clients. Clients not configured will not be able to log in to the trading gateway. In order for a client to submit orders, a login request, with the relevant username and password, must first be sent to the trading gateway. Similarly, log-out request must be sent when the client logs out. Thereafter the client can submit orders by publishing order messages to the trading gateway via UDP ports. Traders cannot submit orders that are smaller than the LOB's tick size (controls the smallest order size). The allowable order types that are not conditioned on time are \cite{jse2020trading}: 

\begin{enumerate}
    \item \textbf{Market orders (MO)} -- contains the quantity of shares to trade, but not the price, and executes against each orders in the LOB on the contra side until it is fully filled. Orders submitted during the auction call will remain in the LOB until uncrossing is done. All market orders that are not filled will expire.
    \item \textbf{Limit order (LO)} -- displays the quantity and price which may execute against a trade or expire based on it's time-in-force (TIF).
    \item \textbf{Hidden orders (HO)} -- allows traders to submit orders which are completely hidden from the market (price and volume). These have a lower priority than all visible active orders. These orders can execute against other visible and hidden orders. Hidden limit orders that are submitted during an auction call are rejected. The quantity of a hidden order must meet the minimum reserve size (MRS)\footnote{The minimum order quantity for orders to qualify as hidden limit orders.}. Each hidden order also has a minimum execution size (MES)\footnote{The minimum quantity of the hidden limit order which is permitted to execute}. After a trade executes against a hidden order, if the remaining quantity is less than the MES (> MRS) then the order will expire. If the remaining quantity is greater than the MRS and MES then the order will only expire when it executes or based on it's TIF (whichever comes first). Hidden orders may only be submitted during the continuous trading session.
    \item \textbf{Stop order and stop limit order (SO \& SL)} -- a stop order is a market order with a stop price. These orders do not enter the order book (remain unelected) until their stop price is reached. When the stop price is reached, the stop order becomes a market order. Similarly, a stop limit order is a limit order with a stop price. These orders do not enter the order book until their stop price is reached. When the stop price is reached, the stop order becomes a limit order. Stop and stop limit buy orders will be elected if the last traded price is equal to or greater than the stop price. Similarly, stop and stop limit sell orders will be elected if the last traded price is equal to or less than the stop price. An incoming stop or stop limit order may be immediately elected on receipt if the stop price has already been reached. These orders will also only be elected at the end of the execution of an order.
\end{enumerate}

Some of the above orders can also have a time-in-force (TIF) which cannot be amended after the order is placed. The following time in force options are available \cite{jse2020trading}:

\begin{enumerate}
    \item \textbf{At the opening (OPG)} -- only accepted during the opening auction and are used to direct orders only towards this session. OPG orders that are not filled during the uncrossing will expire at the end of the opening auction. If there is no opening auction scheduled, OPG orders will be rejected.
    \item \textbf{Good for auction (GFA)} -- orders of this type that are submitted will be parked\footnote{While orders are parked they cannot be executed.} until the next auction and are injected at the start of the auction. GFA orders that are not filled during the uncrossing will be parked for the next auction and are only removed when they are filled or cancelled. If there is no auction session scheduled, GFA orders will be rejected.
    \item \textbf{Good for intraday auction (GFX)} -- These orders are parked until the next intraday auction and are injected at the start of the auction. As opposed to GFA orders, GFX orders that are not filled during uncrossing will expire at the end of the intraday auction. If there is no intraday auction scheduled, GFX orders will be rejected.
    \item \textbf{At the close (ATC)} -- submitted orders of this type are parked until the next closing auction. They are injected at the start of the auction and if they are not filled, will expire at the end of the closing auction. If there is no closing auction scheduled, ATC orders will be rejected.
    \item \textbf{Day (DAY)} -- orders that expire at the end of the trading day. If an order does not specify a TIF, it will default to DAY.
    \item \textbf{Immediate or cancel (IOC)} -- can be partially or fully filled. An IOC that is only partially filled will cancel immediately after execution. IOC orders are rejected during auction calls.
    \item \textbf{Fill or kill (FOK)} -- this order type differs from IOC orders in that they are either fully filled or expired. Partially filled orders are not allowed.
    \item \textbf{Good till cancel (GTC)} -- can remain in the order book for a maximum of 90 calendar days or until the order is filled or canceled.
    \item \textbf{Good till date (GTD)} -- as opposed to GTC orders, these orders remain in the order book until the order is filled, cancelled or a specified expiration date is reached (maximum active time is 90 days). GTD orders do not allow for the specification of an expiration time (only date).
    \item \textbf{Good till time (GTT)} -- remain in the order book until the specified expiry time is reached in the trading day. GTT orders that have an expiry time during an auction will not expire until uncrossing is finished.
    \item \textbf{Closing price cross (CPX)} -- these orders are parked until the start of the closing price cross session. Unexecuted CPX orders are expired at the end of the closing price cross session. Stop and stop limit orders are not allowed to have a TIF of type CPX.
\end{enumerate}

Valid combinations of order types and TIF are shown in Table \ref{table:tif order combinations} below.

\begin{table}[H]
    \small
    \centering
    \begin{tabular}{lc|c|c|c|c} \toprule
        {TIF} & {Market} & {Limit} & {Hidden} & {Stop} & {Stop} \\
        {} & {} & {} & {limit} & {} & {limit} \\ \midrule
        IOC & \cellcolor{tablegreen} & \cellcolor{tablegreen} & \cellcolor{tablegreen} & \cellcolor{tablegreen} & \cellcolor{tablegreen} \\ \hline
        FOK & \cellcolor{tablegreen} & \cellcolor{tablegreen} & \cellcolor{tablegreen} & \cellcolor{tablegreen} & \cellcolor{tablegreen} \\ \hline
        DAY & \cellcolor{tablegreen} & \cellcolor{tablegreen} & \cellcolor{tablegreen} & \cellcolor{tablegreen} & \cellcolor{tablegreen} \\ \hline
        GFA & \cellcolor{tablegreen} & \cellcolor{tablegreen} & \cellcolor{tablegreen} & \cellcolor{tablegreen} & \cellcolor{tablegreen} \\ \hline
        GFX & \cellcolor{tablegreen} & \cellcolor{tablegreen} & \cellcolor{tablered} & \cellcolor{tablered} & \cellcolor{tablered} \\ \hline
        OPG & \cellcolor{tablegreen} & \cellcolor{tablegreen} & \cellcolor{tablered} & \cellcolor{tablered} & \cellcolor{tablered} \\ \hline
        ATC & \cellcolor{tablegreen} & \cellcolor{tablegreen} & \cellcolor{tablered} & \cellcolor{tablered} & \cellcolor{tablered} \\ \hline
        GTC & \cellcolor{tablegreen} & \cellcolor{tablegreen} & \cellcolor{tablegreen} & \cellcolor{tablegreen} & \cellcolor{tablegreen} \\ \hline
        GTD & \cellcolor{tablegreen} & \cellcolor{tablegreen} & \cellcolor{tablegreen} & \cellcolor{tablegreen} & \cellcolor{tablegreen} \\ \hline
        GTT & \cellcolor{tablegreen} & \cellcolor{tablegreen} & \cellcolor{tablegreen} & \cellcolor{tablegreen} & \cellcolor{tablegreen} \\ \hline
        CPX & \cellcolor{tablegreen} & \cellcolor{tablegreen} & \cellcolor{tablegreen} & \cellcolor{tablegreen} & \cellcolor{tablegreen} \\ \bottomrule
    \end{tabular} \\
    \begin{tabular}{llll}
        \cellcolor{tablegreen} & Accepted & \cellcolor{tablered} & Rejected
    \end{tabular}
    \caption{Valid order types and time-in-force (TIF) combinations for the submission of new orders to the trading gateway. \label{table:tif order combinations}}
\end{table}

Lastly, CoinTossX also allows for multiple trading session types which correspond to that of the JSE. The trading sessions below and the rules adopted by each are configurable in the \texttt{data/tradingSessionsCron.properties} file. The times shown are those of the JSE, provided for context \cite{jse2020trading}. All trading session start and end times can be specified by the user through the cron expressions in the data directory. During the auction call sessions orders are not matched immediately, rather, they are matched at the end of the call auction at a single price using a volume maximizing price-discovery process. This process finds the price that will match the most number or buy and sell orders.

\begin{enumerate}
    \item \textbf{Start of trading} (\emph{07:00 -- 08:30}) -- no orders can be submitted or executed executed during this session. Traders will be able to cancel, but not submit, orders during this session.
    \item \textbf{Opening auction call sessions} (\emph{08:30 -- 09:00})
    \item \textbf{Continuous trading session} (\emph{09:00 -- 16:50}) -- the system will continuously match incoming orders against those in the order book according to the price-time priority execution rule.
    \item \textbf{Volatility auction call session} (\emph{triggered}) -- this session will only trigger when a stock's circuit breaker tolerance level has been breached. volatility auction call sessions last for a scheduled period of 5 minutes. The orders accumulated during this session will be executed at the uncrossing based on the volume maximizing algorithm.
    \item \textbf{Intraday auction call session} (\emph{12:00 -- 12:15})
    \item \textbf{Closing auction call session} (\emph{16:50 -- 17:00})
    \item \textbf{Closing price publication} session (\emph{17:00 -- 17:05}) -- no orders can be submitted or executed executed during this session. Traders will be able to cancel, but not submit, orders during this session.
    \item \textbf{Closing price cross session} (\emph{17:05 -- 17:10}) -- trading will only take place at the closing price that was published during the closing price publication session.
    \item \textbf{Post close session} (\emph{17:05 -- 18:15}) -- traders will be able to cancel, but not submit, orders during this session.
    \item \textbf{Halt} (\emph{manually envoked}) -- A halt session may be activated by the user during which time no order equests may be executed. Traders will however be able to cancel orders.
    \item \textbf{Halt and close} (\emph{manually envoked}) -- The behaviour is the same as the halt session except closing price calculations will be performed.
    \item \textbf{Pause} (\emph{manually envoked}) -- No executions will take place during the pause session. Traders will be able to submit, amend or cancel orders during this session, however, market orders will expire at the end of the session.
    \item \textbf{Re-opening auction call} (\emph{manually envoked}) -- The user may manually invoke the re-opening auction call session when resuming from a manual trading halt or a trading pause.
    \item \textbf{Trade reporting} (\emph{08:00 -- 18:15}) -- trades and statistics, where applicable, will be published through the market data gateways.
\end{enumerate}

\begin{table*}[!htb]
    \renewcommand{\arraystretch}{2}
    \scriptsize
    \centering
    \begin{tabular}{lc|c|c|c|c|c|c|c|c|c|c|c|c|c|c} \toprule
        {} & \multicolumn{11}{c}{Order type/time-in-force} & \multicolumn{4}{c}{Order type/time-in-force} \\ \cmidrule(lr){2-12} \cmidrule(lr){13-16}
        {Session} & {OPG} & {ATC} & {IOC} & {FOK} & {GTC} & {GTD} & {GTT} & {GFA} & {GFX} & {DAY} & {CPX} & {MO} & {LO} & {\shortstack{SO \&\\SL}} & {HL} \\ \midrule
        \shortstack{Start of\\trading} & \cellcolor{tablered} & \cellcolor{tablered} & \cellcolor{tablered} & \cellcolor{tablered} & \cellcolor{tablepurple} & \cellcolor{tablepurple} & \cellcolor{tablered} & \cellcolor{tablered} & \cellcolor{tablered} & \cellcolor{tablered} & \cellcolor{tablered} & \cellcolor{tablered} & \cellcolor{tablered} & \cellcolor{tablered} & \cellcolor{tablered} \\ \hline
        \shortstack{Opening\\auction call} & \cellcolor{tablegreen} & \cellcolor{tableyellow} & \cellcolor{tablered} & \cellcolor{tablered} & \cellcolor{tablegreen} & \cellcolor{tablegreen} & \cellcolor{tablegreen} & \cellcolor{tablegreen} & \cellcolor{tableyellow} & \cellcolor{tablegreen} & \cellcolor{tableyellow} & \cellcolor{tablegreen} & \cellcolor{tablegreen} & \cellcolor{tableyellow} & \cellcolor{tablered} \\ \hline
        \shortstack{Continuous\\trading} & \cellcolor{tablered} & \cellcolor{tableyellow} & \cellcolor{tableblue} & \cellcolor{tableblue} & \cellcolor{tablegreen} & \cellcolor{tablegreen} & \cellcolor{tablegreen} & \cellcolor{tableyellow} & \cellcolor{tableyellow} & \cellcolor{tablegreen} & \cellcolor{tableyellow} & \cellcolor{tableblue} & \cellcolor{tablegreen} & \cellcolor{tableyellow} & \cellcolor{tablegreen} \\ \hline
        \shortstack{Volatility\\auction call} & \cellcolor{tablered} & \cellcolor{tableyellow} & \cellcolor{tablered} & \cellcolor{tablered} & \cellcolor{tablegreen} & \cellcolor{tablegreen} & \cellcolor{tablegreen} & \cellcolor{tablegreen} & \cellcolor{tableyellow} & \cellcolor{tablegreen} & \cellcolor{tableyellow} & \cellcolor{tablegreen} & \cellcolor{tablegreen} & \cellcolor{tableyellow} & \cellcolor{tablered} \\ \hline
        \shortstack{Intraday\\auction call} & \cellcolor{tablered} & \cellcolor{tableyellow} & \cellcolor{tablered} & \cellcolor{tablered} & \cellcolor{tablegreen} & \cellcolor{tablegreen} & \cellcolor{tablegreen} & \cellcolor{tablegreen} & \cellcolor{tablegreen} & \cellcolor{tablegreen} & \cellcolor{tableyellow} & \cellcolor{tablegreen} & \cellcolor{tablegreen} & \cellcolor{tableyellow} & \cellcolor{tablered} \\ \hline
        \shortstack{Closing\\auction call} & \cellcolor{tablered} & \cellcolor{tablegreen} & \cellcolor{tablered} & \cellcolor{tablered} & \cellcolor{tablegreen} & \cellcolor{tablegreen} & \cellcolor{tablegreen} & \cellcolor{tablegreen} & \cellcolor{tablered} & \cellcolor{tablegreen} & \cellcolor{tableyellow} & \cellcolor{tablegreen} & \cellcolor{tablegreen} & \cellcolor{tableyellow} & \cellcolor{tablered} \\ \hline
        \shortstack{Closing price\\publication} & \cellcolor{tablered} & \cellcolor{tablered} & \cellcolor{tablered} & \cellcolor{tablered} & \cellcolor{tableyellow} & \cellcolor{tableyellow} & \cellcolor{tableyellow} & \cellcolor{tablered} & \cellcolor{tablered} & \cellcolor{tableyellow} & \cellcolor{tableyellow} & \cellcolor{tableyellow} & \cellcolor{tableyellow} & \cellcolor{tablered} & \cellcolor{tablered} \\ \hline
        \shortstack{Closing price\\cross session} & \cellcolor{tablered} & \cellcolor{tablered} & \cellcolor{tablered} & \cellcolor{tablered} & \cellcolor{tablegreen} & \cellcolor{tablegreen} & \cellcolor{tablegreen} & \cellcolor{tablered} & \cellcolor{tablered} & \cellcolor{tablegreen} & \cellcolor{tablegreen} & \cellcolor{tablegreen} & \cellcolor{tablegreen} & \cellcolor{tablered} & \cellcolor{tablered} \\ \hline
        Post close & \cellcolor{tablered} & \cellcolor{tablered} & \cellcolor{tablered} & \cellcolor{tablered} & \cellcolor{tablered} & \cellcolor{tablered} & \cellcolor{tablered} & \cellcolor{tablered} & \cellcolor{tablered} & \cellcolor{tablered} & \cellcolor{tablered} & \cellcolor{tablered} & \cellcolor{tablered} & \cellcolor{tablered} & \cellcolor{tablered} \\ \hline
        Halt & \cellcolor{tablered} & \cellcolor{tablered} & \cellcolor{tablered} & \cellcolor{tablered} & \cellcolor{tablered} & \cellcolor{tablered} & \cellcolor{tablered} & \cellcolor{tablered} & \cellcolor{tablered} & \cellcolor{tablered} & \cellcolor{tablered} & \cellcolor{tablered} & \cellcolor{tablered} & \cellcolor{tablered} & \cellcolor{tablered} \\ \hline
        \shortstack{Halt and\\close} & \cellcolor{tablered} & \cellcolor{tablered} & \cellcolor{tablered} & \cellcolor{tablered} & \cellcolor{tablered} & \cellcolor{tablered} & \cellcolor{tablered} & \cellcolor{tablered} & \cellcolor{tablered} & \cellcolor{tablered} & \cellcolor{tablered} & \cellcolor{tablered} & \cellcolor{tablered} & \cellcolor{tablered} & \cellcolor{tablered} \\ \hline
        Pause & \cellcolor{tablered} & \cellcolor{tableyellow} & \cellcolor{tablered} & \cellcolor{tablered} & \cellcolor{tablegreen} & \cellcolor{tablegreen} & \cellcolor{tablegreen} & \cellcolor{tableyellow} & \cellcolor{tableyellow} & \cellcolor{tablegreen} & \cellcolor{tableyellow} & \cellcolor{tablegreen} & \cellcolor{tablegreen} & \cellcolor{tableyellow} & \cellcolor{tablered} \\ \hline
        \shortstack{Re-opening\\auction call} & \cellcolor{tablered} & \cellcolor{tableyellow} & \cellcolor{tablered} & \cellcolor{tablered} & \cellcolor{tablegreen} & \cellcolor{tablegreen} & \cellcolor{tablegreen} & \cellcolor{tablegreen} & \cellcolor{tablegreen} & \cellcolor{tablegreen} & \cellcolor{tableyellow} & \cellcolor{tablegreen} & \cellcolor{tablegreen} & \cellcolor{tableyellow} & \cellcolor{tablered} \\ \hline
        \shortstack{FCO auction\\call session} & \cellcolor{tablered} & \cellcolor{tableyellow} & \cellcolor{tablered} & \cellcolor{tablered} & \cellcolor{tablegreen} & \cellcolor{tablegreen} & \cellcolor{tablegreen} & \cellcolor{tablegreen} & \cellcolor{tablegreen} & \cellcolor{tablegreen} & \cellcolor{tableyellow} & \cellcolor{tablegreen} & \cellcolor{tablegreen} & \cellcolor{tableyellow} & \cellcolor{tablered} \\ \bottomrule
    \end{tabular} \\
    \begin{tabular}{llllllllll}
        \cellcolor{tablegreen} & Accepted & \cellcolor{tablered} & Rejected & \cellcolor{tableyellow} & \shortstack{Accepted and parked\\until injected} & \cellcolor{tableblue} & \shortstack{Accepted and expired immediately if\\they do not execute upon aggression} & \cellcolor{tablepurple} & \shortstack{Carried forward from\\the previous day}
    \end{tabular}
    \caption{Valid trading session and order type/TIF combinations for the submission of new orders to the trading gateway. \label{table:session tif/order combinations}}
\end{table*}

\subsection{Testing framework \label{sec:Testing framework}}
CoinTossX has been successfully deployed locally as well as to remote servers such as \href{https://azure.microsoft.com/en-us/}{Microsoft Azure}, \href{https://www.chpc.ac.za/}{CHPC} and \href{https://www.wits.ac.za/mss/}{TW Kambule Mathematical Sciences Laboratories} provided servers (using 4 re-purposed legacy TACC Ranger blades, see Table \ref{table:hardware}). Deployment to high-performance compute solutions such as \href{http://hpc.uct.ac.za/}{UCT HPC} was found to be infeasible for a number of reasons: First, facilities such as UCT HPC often perform computations by relying on an MPI approach using one of a variety of different job management systems where the submission of ``jobs'' to ``worker nodes'' is highly constrained by the preferences of systems administrators managing many different use cases; the worker nodes are not equivalent to virtual machines, rather they receive tasks from the ``head node'' to be executed in parallel. Simplistically this means that, for example, web interfaces will not be easily accessible from worker nodes due to the system and network architecture - but this more generally impacts any client worker interactions when agents interact via clients through some centralised architecture - here the matching engine; but this can be any interaction landscape. In general, such systems are poorly suited for real-time and reactive use cases. This is because, by design, many high-performance computing facilities, such as \href{http://hpc.uct.ac.za/}{UCT HPC} or \href{https://www.chpc.ac.za/}{CHPC}, are not high-throughput facilities. They are typically not well suited for large scale high-concurrency, low latency market and agent-based simulation problems that necessarily need to be highly reactive \footnote{Reactive is used here in the broad sense of system architectures that require the fast propagation of data changes and relationships within a system with many clients, but require high coherence with low coupling.} in nature. We believe that this is an important design perspective that is often not well considered when designing advanced market (or social science) simulators. 

\subsubsection{Unit testing}

The functionality of the software was tested using the test cases made available online from the JSE. The requirements were taken from \cite{jse2020trading}. Testing of the trading sessions were restricted to the continuous and intraday auction trading sessions using only the DAY TIF. Unit tests were implemented to cover the testing of the functional requirements of the software while individual throughput and latency performance tests were implemented in conjunction with the \href{https://github.com/openjdk/jmh}{Java Microbenchmark Harness}\footnote{JMH is a Java harness for building, running, and analysing nano/micro/milli/macro benchmarks written in Java and other languages targetting the JVM} (JMH) to test methods whose performance were critical. These tests, covering the majority of the application, provide a safety net to allow changes to be made to the code without breaking existing functionality.

The outputs of these tests were not compared to the JSE or another exchange as the data is not available by the industry. The JSE’s test environment is also closed and does not provide realistic order-book dynamics. This paper focuses mainly on the results of latency and throughput tests, however, the types and results of the extensive list of tests performed can be found by referring to \citet{sing2017jse}.

\subsubsection{Order-flow testing}

Matching engine integrity was evaluated using unit tests and, instead of an agent-based approach, simulations and tests aim to understand throughput and latency were carried out with a  8-variate marked Hawkes process \cite{hawkes1971spectra, ogata1988statistical}. This provides a flexible framework to simulate a market data feed with varying throughput, with full control over the trade and quote conditional intensities. The software and Hawkes client processes were deployed on one server which affected the performance of all components. The mutually exciting processes correspond to 8 different order types that are considered in the testing framework. The testing framework only considers basic aggressive and passive market and limit orders as in \citet{large2007measuring}:

\begin{table}[H]
    \setlength{\tabcolsep}{2pt}
    \small
    \centering
    \begin{tabular}{lccccl} \toprule
        {Type} & {Event} & {Ask or} & {Immediate} & {Moves} & {Descritpion} \\
        {\#} & {type} & {Bid} & {execution} & {prices} & {} \\ \midrule
        1 & Trade & Ask & \ding{51} \ (MO) & \ding{51} & \shortstack{Market buy that\\moves the ask} \\
        2 & Trade & Bid & \ding{51} \ (MO) & \ding{51} & \shortstack{Market sell that\\moves the bid} \\
        3 & Add & Bid & \ding{55} \ (LO) & \ding{51} & Bid between quotes \\
        4 & Add & Ask & \ding{55} \ (LO) & \ding{51} & Ask between quotes \\
        5 & Trade & Ask & \ding{51} \ (MO) &  \ding{55} & \shortstack{Market buy does\\not move ask} \\
        6 & Trade & Bid & \ding{51} \ (MO) &  \ding{55} & \shortstack{Market sell does\\not move bid} \\
        7 & Add & Bid & \ding{55} \ (LO) &  \ding{55} & \shortstack{Bid at or below\\best bid} \\
        8 & Add & Ask & \ding{55} \ (LO) &  \ding{55} & \shortstack{Ask at or above\\best ask} \\ \bottomrule
    \end{tabular}
    \caption{Order event types used for the 8-variate Hawkes process in the testing framework simulation following the approach of \citet{large2007measuring}. \label{table:order types}}
\end{table}

The simulation is conducted using the intensity-based thinning algorithm as introduced by \citet{lewis1979simulation} and modified by \citet{ogata1981lewis} and is used to define the time at which orders arrive. Each thinning algorithm submitting large numbers of orders may be thought of as clients. So, in the case of the testing framework, a single client is associated with each stock and is meant to represent a large group of traders investing in that stock. The prices and volumes are generated based on the order type in a fairly random manner.

First, a maximum LOB depth $M = 10$ is specified. A lower limit for the price at which buy limit orders are generated is set to $L_b = 25000$. Similarly, an upper limit for the price at which sell limit orders are generated is set to $H_s = 25057$. The LOB is initialised with an initial limit and buy $I_b = 25034$ and sell $I_s = 25057$ order. Thereafter prices and volumes are generated according to a random normal distribution in these bounds. According to this bids and asks can cross each other - which is not realistic. Nonetheless it's purpose is merely to demonstrate the applications ability. The VWAP is used to calculate the price for an aggressive buy/sell trade (not the same as the execution price) that executes against the LOB. That is, if an aggressive trade affects the first $k$ highest/lowest levels of the order book then the price is calculated as:
\begin{equation*}
    \frac{\sum_{i = 1}^{k} \mbox{Price}_i \times \mbox{Volume}_i}{\sum_{i = 1}^{k} \mbox{Volume}_i}
\end{equation*}
For the next section, test scenarios were created to test the performance of the software by evaluating the impact of multiple clients-stock pairs. That is, each stock is assigned a unique client who sends high volumes of orders. By increasing the number of stocks and clients after each subsequent run, the volume of messages being processed increases as well - during which time the throughput per second and latency are automatically recorded. These performance tests have been conducted on multiple computers, all with different operating systems and hardware specifications. Here the performance results are documented by running the application on the two machines listed in Table \ref{table:hardware}. The average start-up time of the web application is approximately 50 seconds.

\begin{table}[H]
    \small
    \centering
    \begin{tabular}{>{\centering\arraybackslash}m{1cm}>{\centering\arraybackslash}m{2cm}>{\centering\arraybackslash}m{1cm}>{\centering\arraybackslash}m{3cm}} \toprule
        {Machine} & {\shortstack{Operating\\System}} & {Memory} & {Processors} \\ \midrule
        WITS MSS Server & 64-bit Linux Ubuntu 16.04-LTS & $32$GB & \shortstack{$4 \times$ Quadcore AMD \\ Opteron 8356 \\@ 2.3GHz (16 cores)} \\
        \shortstack{Standard\\A4m V2\\Azure VM} & \shortstack{64-bit Linux\\Ubuntu\\ 18.04-LTS} & 32GB & \shortstack{$4 \times$ Intel Xeon\\CPU E5-2673 v3\\@ 2.4GHz (32 cores)} \\ \bottomrule
    \end{tabular}
    \caption{Hardware specifications of machines used for throughput and latency testing used to compare the cloud solution (Azure) to the dedicated physical hardware solution (WITS MSS Server blade) \label{table:hardware}.}
\end{table}

Although not presented here, limit order book storage testing was also conducted on the WITS Mathematical Science Support provided server hardware and demonstrated the ability to store thousands of orders at each price point. The design of the LOB also allows orders to be added and removed easily

\subsubsection{Latency tests}

\begin{figure*}[!htb]
    \subfloat[Wits Server high specification machine \label{figa:osx}]{\includegraphics[width=.49\textwidth]{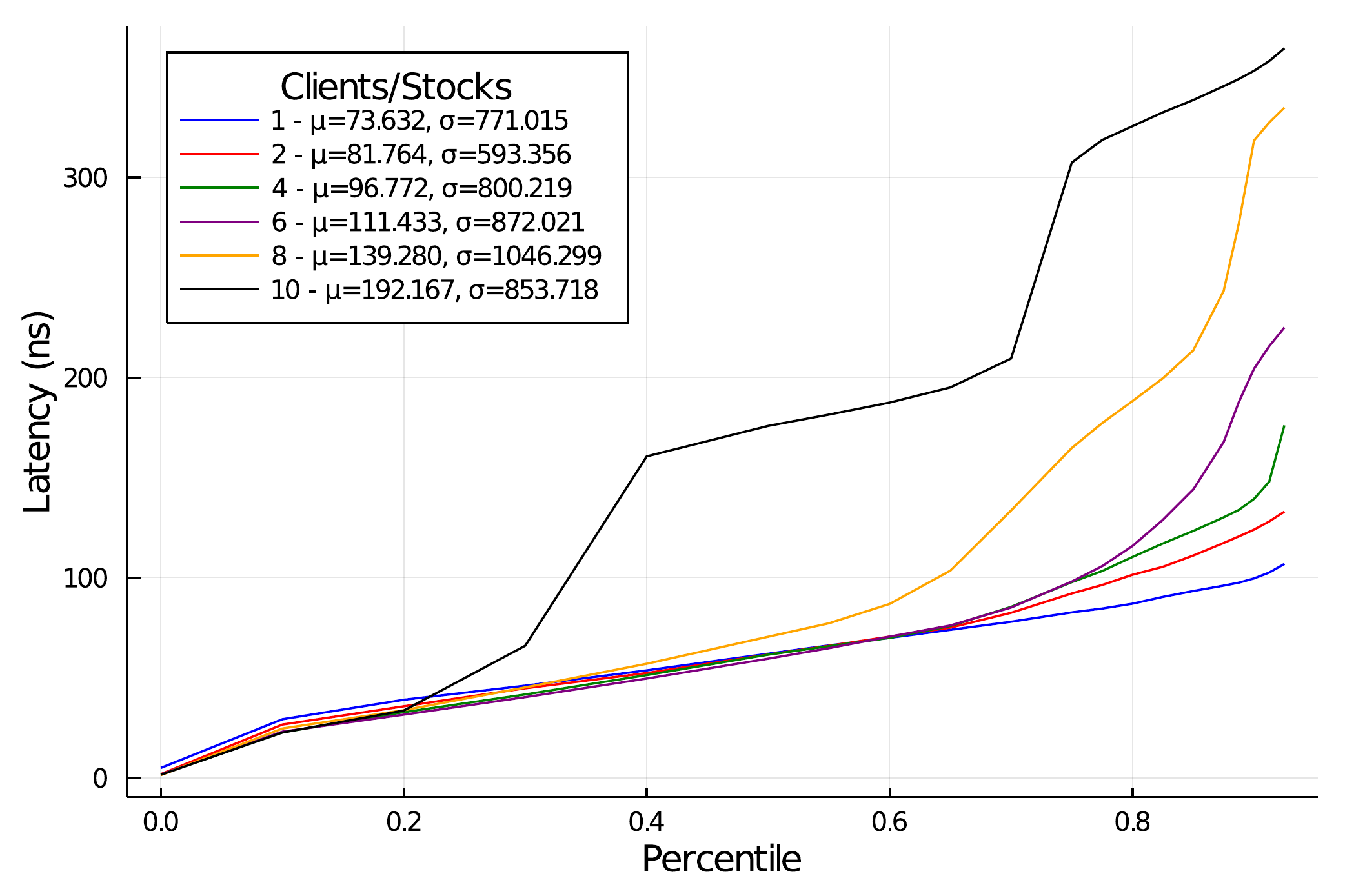}}
    \subfloat[Azure medium specification virtual machine \label{figc:linux}]{\includegraphics[width=.49\textwidth]{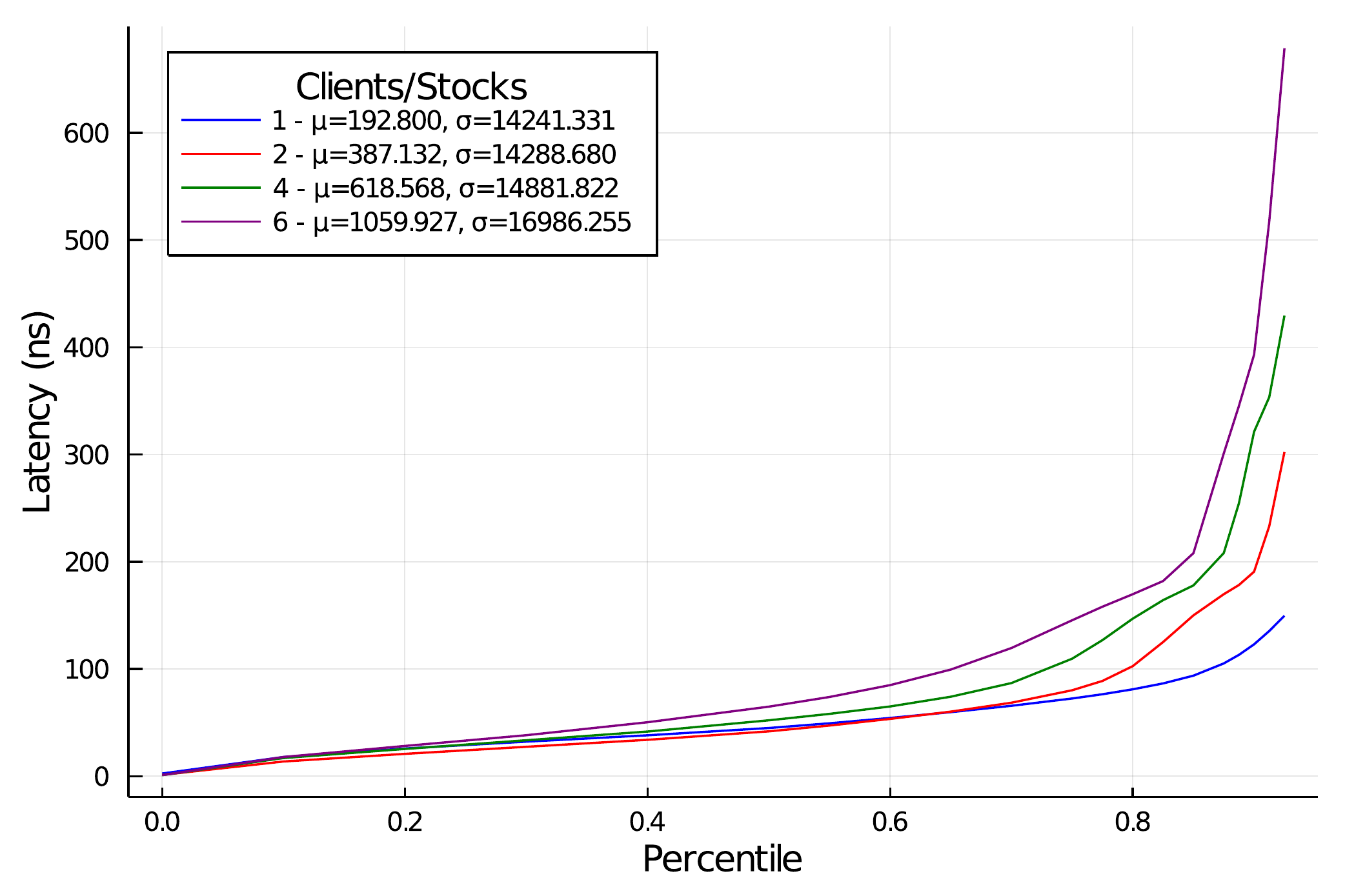}}
    \caption{A High Dynamic Range (HDR) Histogram graph showing the latency percentiles of the matching engine for increasing numbers of stocks and clients. In the testing framework each client submits approximately 110000 market/limit orders. \label{fig:latency}}
\end{figure*}

The latency was tested and visualized using the \href{https://github.com/HdrHistogram/HdrHistogram}{HdrHistogram} library \footnote{HdrHistogram is designed for recording histograms of value measurements in latency and performance sensitive applications. Measurements show value recording times as low as 3-6 nanoseconds on modern (circa 2012) Intel CPUs}. With every run, each client submits approximately 110 000 orders. The time it takes to process all these orders is then measured and compared between machines. To reproduce Figure \ref{fig:latency} simply re-run the Hawkes simulations for the system to write latency and throughput results to file.

Figure \ref{fig:latency} shows the latency results, in nanoseconds, for runs with differing numbers of active client-stock pairs on each machine. The latency increases as the number of clients-stock pairs increase - since each client is associated with multiple threads running processes in parallel. Due to there being only 4 CPUs on the Azure VM, running more than 6 clients simultaneously was found to be infeasible. For this reason the user should ensure that the hardware on the server is capable of supporting the desired number of clients. For the high spec Wits Server machine the minimum and maximum latency (measured up to 10 clients) at the 90th percentile is 106ns and 248ns, respectively. Similarly, for the medium spec machine (with a maximum of 6 clients) the minimum and maximum latency at the 90th percentile is 123ns and 393ns, respectively. Therefore, on a high end machine one can expect sub 250ns latency (with 10 clients), while with a medium spec machine one can expect sub 400ns latency (with 6 clinets) \footnote{JSE's average round-trip colocation network latency is sub 100 microseconds and it's matching engine has a latency of 50 microseconds \cite{jse2020latency}.}.

\begin{figure}[H]
    \includegraphics[width=\linewidth]{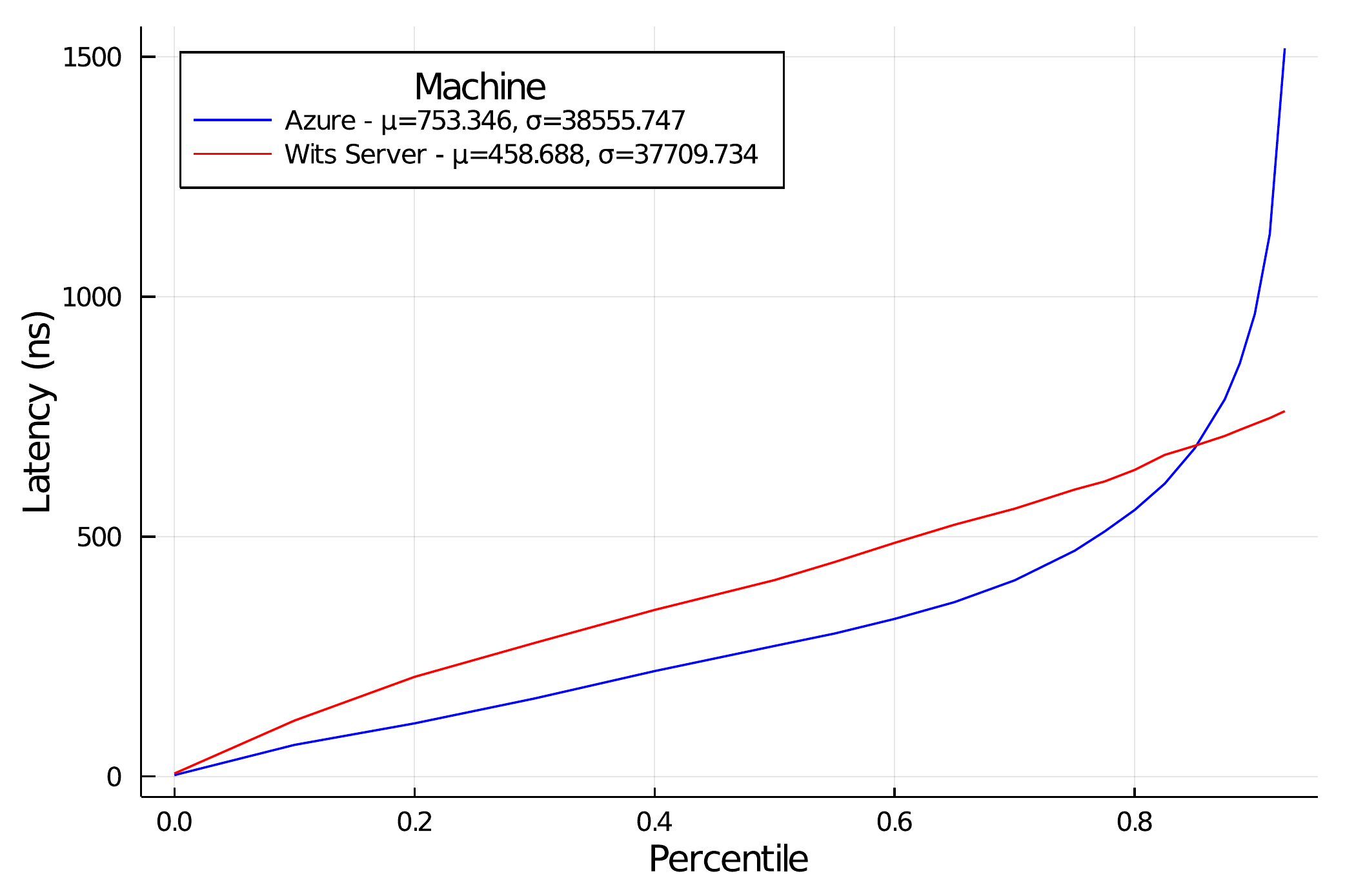}
    \caption{A comparison of latencies during a high volume scenario. One million orders were submitted by a single client on the Wits server and Azure machines. \label{fig:high volume}}
\end{figure}

Figure \ref{fig:high volume} considers the scenario where high volumes of orders are submitted to the trading gateway. One million orders are submitted from a single client and compared across the two machines. For the Wits server, the latency is 735ns at the 90th percentile but maintains a significantly lower latency on average. On the other hand, the Azure server has higher latency's on average with a latency of 964ns at the 90th percentile.

\subsubsection{Throughput tests}
Table \ref{table:throughput} shows the throughput per second as the number of clients-stock pairs increase on each machine. To reproduce the data in Table \ref{table:throughput} simply re-run the Hawkes simulations for the system to write latency and throughput results to file. It should be noted that each client waits for market data updates before calculating the next order to send. The client also waits a few nanoseconds as part of the Hawkes simulation. These delays reduced the throughput of orders sent. The results show that as the number of clients-stock pairs increase, the throughput decreases. The most significant decrease in throughput is when more than 4 clients and stocks are used. Beyond 6 clients the hardware configuration of the Azure VM was found to be insufficient. In total, the high spec machine was shown to be capable of processing more than one million orders in less than a 19 minute simulation period with significantly low latencies. On the other hand, the medium spec machine was capable of processing 450000 and 560000 orders in approximately 44 minute and 2 hour simulation periods for 4 and 6 client-stock pairs, respectively.

\begin{table*}[!htb]
    \small
    \centering
    \begin{tabular}{lcccccc} \toprule
        {} & \multicolumn{3}{c}{Wits Server} & \multicolumn{3}{c}{Microsoft Azure} \\ \cmidrule(lr){2-4} \cmidrule(lr){5-7}
        {\shortstack{client-stock\\pairs}} & {Duration} & {\shortstack{Number of\\orders}} & {\shortstack{Throughput per\\second}} & {Duration} & {\shortstack{Number of\\orders}} & {\shortstack{Throughput per\\second}} \\ \midrule
        1 & 00:00:48.820 & 111646 & 2287 & 00:06:44.917 & 110825 & 274 \\
        2 & 00:04:23.981 & 224562 & 850 & 00:19:38.576 & 222390 & 189 \\
        4 & 00:12:02.204 & 448774 & 621 & 00:43:55.734 & 447878 & 170 \\
        6 & 00:15:12.386 & 669331 & 733 & 02:01:27.494 & 564070 & 77 \\
        8 & 00:20:27.010 & 895080 & 729 & - & - & - \\
        10 & 00:18:52.289 & 1120514 & 989 & - & - & - \\ \bottomrule
    \end{tabular}
    \caption{Tabulated measurements of throughput per second with increasing numbers of client-stock pairs. Based on the Hawkes process parameters each client submits approximately 110000 orders. After each simulation the start and end times are published and used to calculate the throughput per second. The hardware configuration on the Azure VM was found to be insufficient when defining more that 6 clients. All data generated in this table can be acquired at the end of each simulation in the data directory of the start-up folder. \label{table:throughput}}
\end{table*}

\section{Simulation results \label{sec:Simulation results}}
The results in this section relate to the simulation for a single stock using a simple Hawkes process as a proof of concept. Simulation results are contained in three seperate \texttt{.csv} files for each security: one for market orders only, another for limit orders and the last for a snapshot of the limit order book at then end of the simulation. This data is only printed to file at the end of the simulation once all logged-in clients have logged out. The format of these files are shown in snippets \ref{verb:lob file} and \ref{verb:trades file} below. The submission times are given by the UTC standard. \texttt{TradId}'s correspond to the \texttt{OrderId}'s of the limit orders order against which they were executed.

\begin{figure}[H]
    \begin{Verbatim}[fontsize=\scriptsize]
        SecurityId,"OrderId","SubmittedTime","Price","Volume","Side"
        1,"1","2020-11-22 07:43:08.231","25034","1200","Buy"
        1,"2","2020-11-22 07:43:08.696","25057","1000","Sell"
        1,"3","2020-11-22 07:43:11.683","25056","3600","Buy"
        1,"4","2020-11-22 07:43:11.763","25050","2600","Buy"
        1,"5","2020-11-22 07:43:11.915","25048","1200","Sell"
    \end{Verbatim}
    \caption{File format for the output of simulated limit orders sent to the trading gateway. This data is written to file after the simulation is complete and the client ends their session. \label{verb:lob file}}
\end{figure}

\begin{figure}[H]
    \begin{Verbatim}[fontsize=\scriptsize]
        TradeId,"Price","Quantity","CreationTime"
        1,"25056","1200","2020-11-22 07:43:12.352"
        2,"25056","1700","2020-11-22 07:43:12.905"
        3,"25056","700","2020-11-22 07:43:12.906"
        5,"25057","100","2020-11-22 07:43:12.984"
        6,"25057","300","2020-11-22 07:43:13.115"
    \end{Verbatim}
    \caption{File format for the output of simulated market orders sent to the trading gateway. \label{verb:trades file}}
\end{figure}

\begin{figure}[H]
    \centering
    \includegraphics[width=0.48\textwidth]{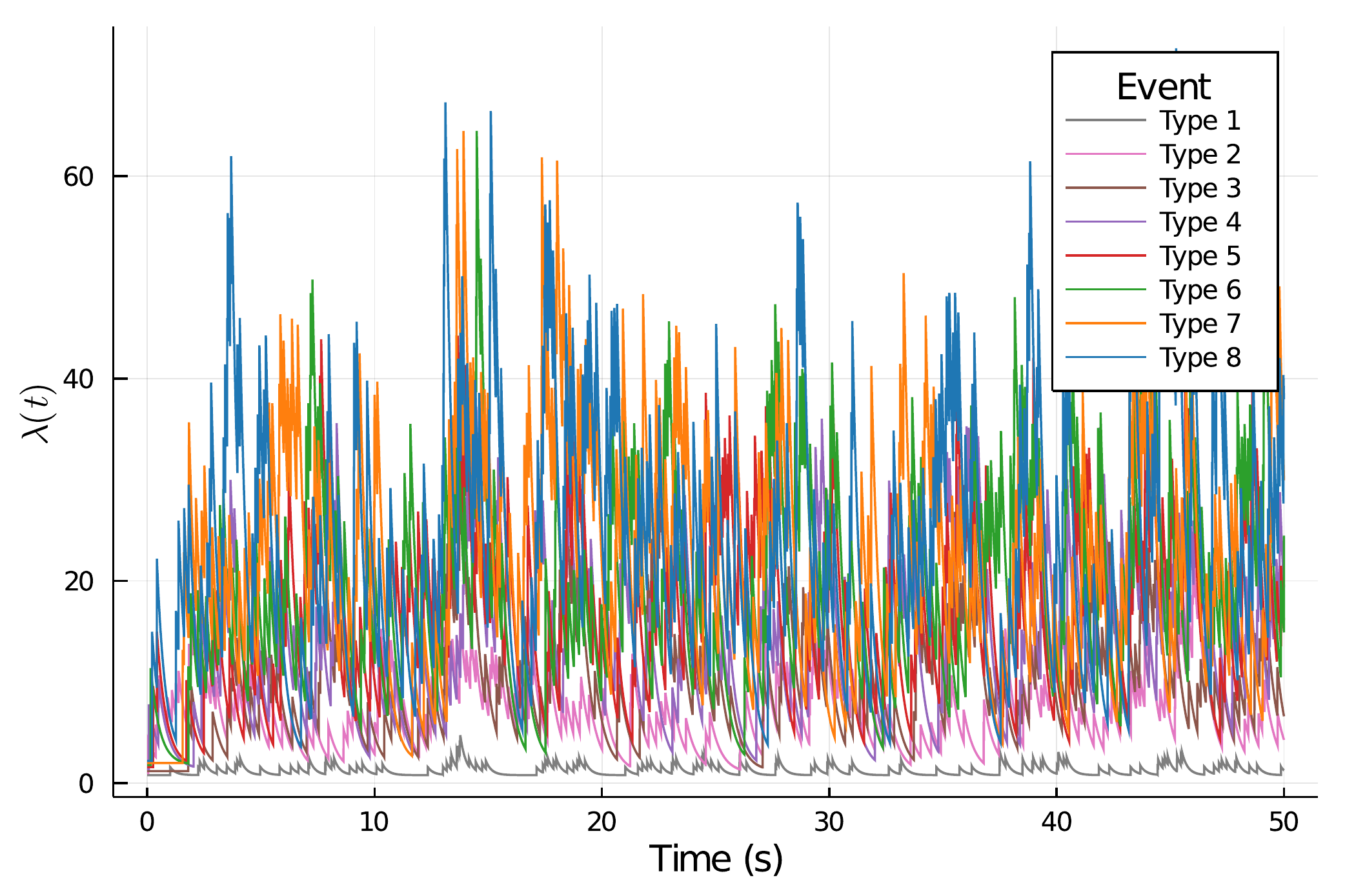}
    \caption{A Graph of the intensities of the 8-variate Hawkes process for 50 seconds of the Hawkes simulation. \label{fig:hawkes intensities}}
\end{figure}

\begin{figure}[H]
    \centering
    \includegraphics[width=0.48\textwidth]{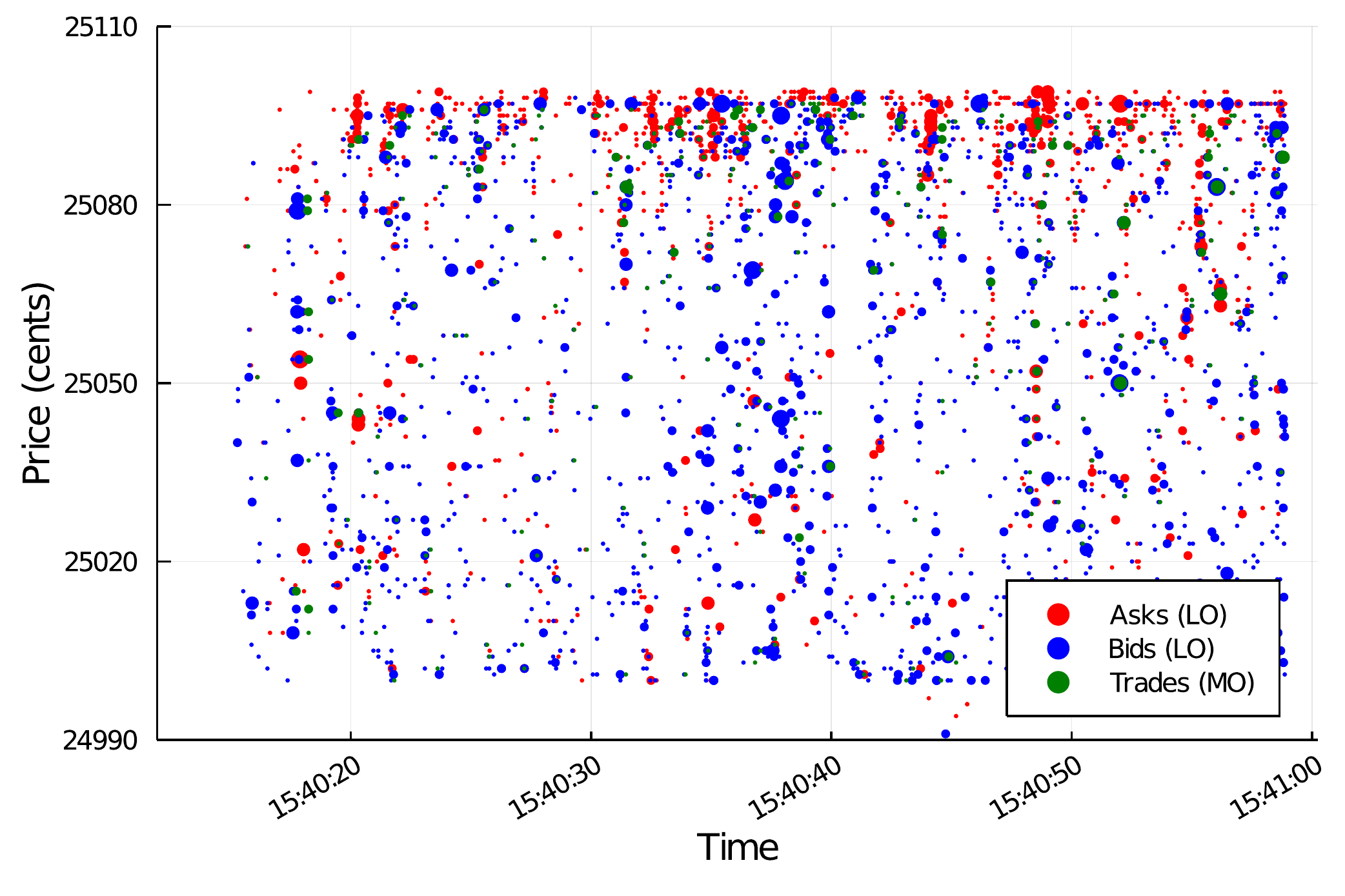}
    \caption{A Graph of the limit and market orders submitted to the trading gateway. As a proof of concept, the Hawkes simulation assigns volumes and prices in a fairly random and unrealistic manner as the bid and ask are allowed to cross (negative spread). The size of the points are proportional to the volume of the order. \label{fig:bubble}}
\end{figure}

\section{Conclusion \label{sec:Conclusion}}
CoinTossX is a low latency high throughput stock exchange. It is configurable and allows users to view the limit order book in real time as well as for multiple clients to connect and send orders to the exchange. The exchange supports multiple stocks and a variety of different trading session. Here Hawkes processes are used to provide a simple but robust simulation of order arrivals for infrastructure testing. The CoinTossX website can be enhanced to analyze the data that is processed. The exchange provides a realistic platform for agent based models exploration. The software was designed so that different matching logic algorithms are in separate Java classes. This allows the logic to be changed to test any variations of the matching logic or market rules. Hence, additional work can be done to change the matching rules on the engine to test the impact of rules and regulation changes on the limit order book and its dynamics. Since the components are de-coupled, their implementation language can be continually and easily changed to support the latest frameworks \cite{sing2017cointossx}. Future work will be aimed at better understanding the interaction dynamics from increasingly realistic approaches to point-process based simulations {\it e.g.} with trading clients based on Hawkes processes using models such as those of \citet{bacry2014hawkes} and \citet{zheng2014modelling} to investigate the interactions of limit-order and market-order trading agents through a realistic order matching process. Ultimately this type of work is aimed at better understanding high-concurrency agent-based perspectives for market modeling and their impact on model calibration \cite{platt2018can} and causation.

\section{Acknowledgement}

We thank Turgay Celik and Brian Maistry  \href{https://www.wits.ac.za/mss/}{WITS Mathematical Science Support} and the TW  Kambule  Mathematical  Sciences  Laboratories for allowing us to use 4 former TACC Ranger blades for some of the simulation work. We thank Patrick Chang, Dieter Hendricks and Diane Wilcox for various discussions and advice with regards to the project. 
\bibliographystyle{elsarticle-num-names}
\bibliography{References}

\begin{thebibliography}{27}
\expandafter\ifx\csname natexlab\endcsname\relax\def\natexlab#1{#1}\fi
\providecommand{\url}[1]{\texttt{#1}}
\providecommand{\href}[2]{#2}
\providecommand{\path}[1]{#1}
\providecommand{\DOIprefix}{doi:}
\providecommand{\ArXivprefix}{arXiv:}
\providecommand{\URLprefix}{URL: }
\providecommand{\Pubmedprefix}{pmid:}
\providecommand{\doi}[1]{\href{http://dx.doi.org/#1}{\path{#1}}}
\providecommand{\Pubmed}[1]{\href{pmid:#1}{\path{#1}}}
\providecommand{\bibinfo}[2]{#2}
\ifx\xfnm\relax \def\xfnm[#1]{\unskip,\space#1}\fi
\bibitem[{Chang et~al.(2020)Chang, Pienaar, and Gebbie}]{chang2020epps}
\bibinfo{author}{P.~Chang}, \bibinfo{author}{E.~Pienaar},
  \bibinfo{author}{T.~Gebbie}, \bibinfo{title}{The epps effect under
  alternative sampling schemes}, \bibinfo{year}{2020}. \URLprefix
  \url{https://arxiv.org/abs/2011.11281}.
  \href{http://arxiv.org/abs/2011.11281}{{\tt arXiv:2011.11281}}.
\bibitem[{Platt and Gebbie(2018)}]{platt2018can}
\bibinfo{author}{D.~Platt}, \bibinfo{author}{T.~Gebbie},
\newblock \bibinfo{title}{Can agent-based models probe market microstructure?},
\newblock \bibinfo{journal}{Physica A: Statistical Mechanics and its
  Applications} \bibinfo{volume}{503} (\bibinfo{year}{2018})
  \bibinfo{pages}{1092--1106}. \URLprefix
  \url{http://www.sciencedirect.com/science/article/pii/S0378437118309956}.
  \DOIprefix\doi{10.1016/j.physa.2018.08.055}.
\bibitem[{Goosen and Gebbie(2020)}]{goosen2020calibrating}
\bibinfo{author}{K.~Goosen}, \bibinfo{author}{T.~Gebbie},
  \bibinfo{title}{Calibrating high-frequency trading data to agent-based models
  using approximate bayesian computation}, Master's thesis, University of cape
  town, \bibinfo{year}{2020}. \URLprefix
  \url{https://github.com/KellyGoosen1/hft-abm-smc-abc}.
  \DOIprefix\doi{10.25375/uct.12894005.v1}.
\bibitem[{Addison et~al.(2019)Addison, Andrews, Azad, Bardsley, Bauman, Diaz,
  Didik, Fazliddin, Gromoa, Krish, Prins, Ryan, and Villette}]{addison2019low}
\bibinfo{author}{A.~Addison}, \bibinfo{author}{C.~Andrews},
  \bibinfo{author}{N.~Azad}, \bibinfo{author}{D.~Bardsley},
  \bibinfo{author}{J.~Bauman}, \bibinfo{author}{J.~Diaz},
  \bibinfo{author}{T.~Didik}, \bibinfo{author}{K.~Fazliddin},
  \bibinfo{author}{M.~Gromoa}, \bibinfo{author}{A.~Krish},
  \bibinfo{author}{R.~Prins}, \bibinfo{author}{L.~Ryan},
  \bibinfo{author}{N.~Villette},
\newblock \bibinfo{title}{Low-latency trading in the cloud environment},
\newblock in: \bibinfo{booktitle}{2019 IEEE International Conference on
  Computational Science and Engineering (CSE) and IEEE International Conference
  on Embedded and Ubiquitous Computing (EUC)}, \bibinfo{organization}{IEEE},
  \bibinfo{year}{2019}, pp. \bibinfo{pages}{272--282}. \URLprefix
  \url{https://ieeexplore.ieee.org/document/8919600}.
  \DOIprefix\doi{10.1109/CSE/EUC.2019.00060}.
\bibitem[{Sing(2017)}]{sing2017cointossx}
\bibinfo{author}{D.~Sing}, \bibinfo{title}{Cointossx software},
  \bibinfo{year}{2017}. \URLprefix
  \url{https://github.com/dharmeshsing/CoinTossX}.
  \DOIprefix\doi{10.25375/uct.14069552}.
\bibitem[{Sing and Gebbie(2017)}]{sing2017jse}
\bibinfo{author}{D.~Sing}, \bibinfo{author}{T.~Gebbie}, \bibinfo{title}{JSE
  Matching Engine Simulator}, Master's thesis, University of the Witwatersrand,
  \bibinfo{year}{2017}. \URLprefix
  \url{http://wiredspace.wits.ac.za/handle/10539/25136}.
\bibitem[{Exchange(2020)}]{jse2020bda}
\bibinfo{author}{J.~S. Exchange}, \bibinfo{title}{Broker Deal Accounting
  System}, \bibinfo{year}{2020}. \URLprefix
  \url{https://www.jse.co.za/services/technologies/bda-technical-specifications}.
\bibitem[{Jericevich et~al.(2020)Jericevich, Chang, and
  Gebbie}]{jericevich2020comparing}
\bibinfo{author}{I.~Jericevich}, \bibinfo{author}{P.~Chang},
  \bibinfo{author}{T.~Gebbie}, \bibinfo{title}{Comparing the market
  microstructure between two south african exchanges}, \bibinfo{year}{2020}.
  \URLprefix \url{https://arxiv.org/abs/2011.04367}.
  \href{http://arxiv.org/abs/2011.04367}{{\tt arXiv:2011.04367}}.
\bibitem[{Nair(2015)}]{nair2015agent}
\bibinfo{author}{P.~Nair}, \bibinfo{title}{Agent based modelling of a
  single-stock market on the JSE}, Master's thesis, University of the
  Witwatersrand, \bibinfo{year}{2015}. \URLprefix
  \url{http://wiredspace.wits.ac.za/handle/10539/16840}.
\bibitem[{Thompson et~al.(2011)Thompson, Farley, Barker, Gee, and
  Stewart}]{thompson2011disruptor}
\bibinfo{author}{M.~Thompson}, \bibinfo{author}{D.~Farley},
  \bibinfo{author}{M.~Barker}, \bibinfo{author}{P.~Gee},
  \bibinfo{author}{A.~Stewart}, \bibinfo{title}{Disruptor: High performance
  alternative to bounded queues for exchanging data between concurrent
  threads}, \bibinfo{type}{Technical Report}, LMAX, \bibinfo{year}{2011}.
  \URLprefix \url{https://lmax-exchange.github.io/disruptor/}.
\bibitem[{Wilcox and Gebbie(2014)}]{wilcox2014hierarchical}
\bibinfo{author}{D.~Wilcox}, \bibinfo{author}{T.~Gebbie},
\newblock \bibinfo{title}{Hierarchical causality in financial economics},
\newblock \bibinfo{journal}{SSRN}  (\bibinfo{year}{2014}). \URLprefix
  \url{https://ssrn.com/abstract=2544327}.
  \DOIprefix\doi{10.2139/ssrn.2544327}.
\bibitem[{Lussange et~al.(2018)Lussange, Belianin, Bourgeois-Gironde, and
  Gutkin}]{lussange2018bright}
\bibinfo{author}{J.~Lussange}, \bibinfo{author}{A.~Belianin},
  \bibinfo{author}{S.~Bourgeois-Gironde}, \bibinfo{author}{B.~Gutkin},
\newblock \bibinfo{title}{A bright future for financial agent-based models},
\newblock \bibinfo{journal}{SSRN}  (\bibinfo{year}{2018}). \URLprefix
  \url{https://ssrn.com/abstract=3109904}.
  \DOIprefix\doi{10.2139/ssrn.3109904}.
\bibitem[{Platt(2020)}]{platt2020calibration}
\bibinfo{author}{D.~Platt},
\newblock \bibinfo{title}{A comparison of economic agent-based model
  calibration methods},
\newblock \bibinfo{journal}{Journal of Economic Dynamics and Control}
  \bibinfo{volume}{113} (\bibinfo{year}{2020}) \bibinfo{pages}{103859}.
  \URLprefix
  \url{https://www.sciencedirect.com/science/article/pii/S0165188920300294}.
  \DOIprefix\doi{10.1016/j.jedc.2020.103859}.
\bibitem[{LeBaron(2006)}]{lebaron2006agent}
\bibinfo{author}{B.~LeBaron},
\newblock \bibinfo{title}{Agent-based computational finance},
\newblock \bibinfo{journal}{Handbook of Computational Economics}
  \bibinfo{volume}{2} (\bibinfo{year}{2006}) \bibinfo{pages}{1187--1233}.
\bibitem[{Jericevich et~al.(2021)Jericevich, Sing, and Gebbie}]{Jericevich2021}
\bibinfo{author}{I.~Jericevich}, \bibinfo{author}{D.~Sing},
  \bibinfo{author}{T.~Gebbie},
\newblock \bibinfo{title}{Cointossx},
\newblock \bibinfo{journal}{GitHub}  (\bibinfo{year}{2021}). \URLprefix
  \url{https://zivahub.uct.ac.za/articles/software/CoinTossX/14069552}.
  \DOIprefix\doi{10.25375/uct.14069552.v1}.
\bibitem[{Exchange(2020)}]{jse2020trading}
\bibinfo{author}{J.~S. Exchange}, \bibinfo{title}{Equity Market Trading and
  Information Solution JSE Specification Document Volume 00E --- Trading and
  Information Overview}, \bibinfo{edition}{4} ed., \bibinfo{year}{2020}.
  \URLprefix
  \url{https://www.jse.co.za/content/JSEContractSpecificationItems/Volume\%2000E\%20-\%20Trading\%20and\%20Information\%20Overview\%20for\%20Equity\%20Market.pdf}.
\bibitem[{Exchange(2018)}]{jse2020native}
\bibinfo{author}{J.~S. Exchange}, \bibinfo{title}{Equity Market Trading and
  Information Solution JSE Specification Document Volume 01 --- Native Trading
  Gateway}, \bibinfo{edition}{3.05} ed., \bibinfo{year}{2018}. \URLprefix
  \url{https://www.jse.co.za/content/JSETechnologyDocumentItems/Volume\%2001\%20-\%20Native\%20Trading\%20Gateway.pdf}.
\bibitem[{Exchange(2019)}]{jse2020market}
\bibinfo{author}{J.~S. Exchange}, \bibinfo{title}{Equity Market Trading and
  Information Solution JSE Specification Document Volume 05 --- Market Data
  Gateway}, \bibinfo{edition}{3.09} ed., \bibinfo{year}{2019}. \URLprefix
  \url{https://www.jse.co.za/content/JSETechnologyDocumentItems/Volume\%2005\%20-\%20Market\%20Data\%20Gateway\%20(MITCH\%20-\%20UDP).pdf}.
\bibitem[{Logic(2021)}]{real2021aeron}
\bibinfo{author}{R.~Logic}, \bibinfo{title}{Aeron media driver},
  \bibinfo{year}{2021}. \URLprefix
  \url{https://github.com/real-logic/Aeron/wiki}.
\bibitem[{Hawkes(1971)}]{hawkes1971spectra}
\bibinfo{author}{A.~G. Hawkes},
\newblock \bibinfo{title}{Spectra of some self-exciting and mutually exciting
  point processes},
\newblock \bibinfo{journal}{Biometrika} \bibinfo{volume}{58}
  (\bibinfo{year}{1971}) \bibinfo{pages}{83--90}. \URLprefix
  \url{https://www.jstor.org/stable/2334319}. \DOIprefix\doi{10.2307/2334319}.
\bibitem[{Ogata(1988)}]{ogata1988statistical}
\bibinfo{author}{Y.~Ogata},
\newblock \bibinfo{title}{Statistical models for earthquake occurrences and
  residual analysis for point processes},
\newblock \bibinfo{journal}{Journal of the American Statistical association}
  \bibinfo{volume}{83} (\bibinfo{year}{1988}) \bibinfo{pages}{9--27}.
  \URLprefix
  \url{https://www.tandfonline.com/doi/abs/10.1080/01621459.1988.10478560}.
  \DOIprefix\doi{10.1080/01621459.1988.10478560}.
\bibitem[{Large(2007)}]{large2007measuring}
\bibinfo{author}{J.~Large},
\newblock \bibinfo{title}{Measuring the resiliency of an electronic limit order
  book},
\newblock \bibinfo{journal}{Journal of Financial Markets} \bibinfo{volume}{10}
  (\bibinfo{year}{2007}) \bibinfo{pages}{1--25}. \URLprefix
  \url{http://www.sciencedirect.com/science/article/pii/S1386418106000528}.
  \DOIprefix\doi{10.1016/j.finmar.2006.09.001}.
\bibitem[{Lewis and Shedler(1979)}]{lewis1979simulation}
\bibinfo{author}{P.~W. Lewis}, \bibinfo{author}{G.~S. Shedler},
\newblock \bibinfo{title}{Simulation of nonhomogeneous poisson processes by
  thinning},
\newblock \bibinfo{journal}{Naval research logistics quarterly}
  \bibinfo{volume}{26} (\bibinfo{year}{1979}) \bibinfo{pages}{403--413}.
  \URLprefix \url{https://onlinelibrary.wiley.com/doi/10.1002/nav.3800260304}.
  \DOIprefix\doi{10.1002/nav.3800260304}.
\bibitem[{Ogata(1981)}]{ogata1981lewis}
\bibinfo{author}{Y.~Ogata},
\newblock \bibinfo{title}{On lewis' simulation method for point processes},
\newblock \bibinfo{journal}{IEEE transactions on information theory}
  \bibinfo{volume}{27} (\bibinfo{year}{1981}) \bibinfo{pages}{23--31}.
  \URLprefix \url{https://ieeexplore.ieee.org/document/1056305}.
  \DOIprefix\doi{10.1109/TIT.1981.1056305}.
\bibitem[{Exchange(2015)}]{jse2020latency}
\bibinfo{author}{J.~S. Exchange}, \bibinfo{title}{The lowest-latency connection
  to JSE markets}, \bibinfo{year}{2015}. \URLprefix
  \url{https://www.jse.co.za/content/JSETechnologyDocumentItems/3.\%20JSE\%20Colocation\%20Brochure\%202015.pdf}.
\bibitem[{Bacry and Muzy(2014)}]{bacry2014hawkes}
\bibinfo{author}{E.~Bacry}, \bibinfo{author}{J.-F. Muzy},
\newblock \bibinfo{title}{Hawkes model for price and trades high-frequency
  dynamics},
\newblock \bibinfo{journal}{Quantitative Finance} \bibinfo{volume}{14}
  (\bibinfo{year}{2014}) \bibinfo{pages}{1147--1166}.
  \DOIprefix\doi{10.1080/14697688.2014.897000}.
\bibitem[{Zheng et~al.(2014)Zheng, Roueff, and Abergel}]{zheng2014modelling}
\bibinfo{author}{B.~Zheng}, \bibinfo{author}{F.~Roueff},
  \bibinfo{author}{F.~Abergel},
\newblock \bibinfo{title}{Modelling bid and ask prices using constrained hawkes
  processes: Ergodicity and scaling limit},
\newblock \bibinfo{journal}{SIAM Journal on Financial Mathematics}
  \bibinfo{volume}{5} (\bibinfo{year}{2014}) \bibinfo{pages}{99--136}.
  \DOIprefix\doi{10.1137/130912980}.

\end{thebibliography}

\onecolumn
\appendix

\section{Deployment and usage \label{sec:Deployment}}

\begin{minipage}{.17\textwidth}
\begin{figure}[H]
    \begin{adjustbox}{totalheight=\textheight -10\baselineskip} 
        \begin{forest}
            pic dir tree, where level = 0{}{directory}, 
            [CoinTossX
                [ClientSimulator
                    [build]
                    [src
                        [main
                            [java
                                [client]
                                [example]
                                [hawkes]
                            ]
                            [resources]
                        ]
                        [test/java/hawkes]
                    ]
                    [build.gradle, file]
                ]
                [Common]
                [LimitOrderBook]
                [MarketDataGateway]
                [MatchingEngine]
                [MatchingEngineClient]
                [Messages]
                [NativeGateway]
                [Socket]
                [Web]
                [WebEventListener]
                [data
                    [ClientData.csv, file]
                    [hawkesData.properties, file]
                    [Stock.csv, file]
                    [Trader.csv, file]
                    [tradingSessionsCron.properties, file]
                ]
                [build.gradle, file]
                [gradle/wrapper]
                [gradlew, file]
                [gradlew.bat, file]
                [install.gradle, file]
                [lib]
                [scripts]
                [local.properties, file]
                [deploy\_local.gradle, file]
                [remote.properties, file]
                [deploy\_remote.gradle, file]
                [windows.properties, file]
                [settings.gradle, file]
            ]
        \end{forest}
    \end{adjustbox}
\end{figure}
\end{minipage}
\begin{minipage}{.81\textwidth}
Below are the instructions for deploying CoinTossX to a user's local machine or virtual machine in a cloud environment. These instructions apply to Windows, Linux and OSX operating systems. CoinTossX is a Java web application and is built using the \href{https://gradle.org/}{Gradle} build tool. The user need \underline{not} install Gradle or change the version of Gradle installed on their system as the project uses the Gradle wrapper to download and run the required Gradle version\footnote{The Wrapper is a script that invokes a declared version of Gradle, downloading it beforehand if necessary. This standardizes a project on a given Gradle version and simplifies execution (especially when using an IDE)}. Currently CoinTossX is compatible with Java version 8 and Gradle version 6.7.1. The application can be started from the command line, however, it is recommended that the user make use of a Java IDE such as Eclipse or IntelliJ IDEA to automate the start-up process and simplify deployment.

\textbf{Prerequisites}
\begin{enumerate}
    \item JDK version 8 (see Installation section for more detail)
    \item A computer with 4 or more CPU cores and a sufficient amount of RAM (ideally 4 cores 32 GB but can be less depending on the number of clients and stocks used).
    \item A Java IDE such as Eclipse or IntelliJ IDEA.
    \item The user may be required to run commands using Command Prompt (Windows) or Bash (Linux and OSX).
\end{enumerate}

\textbf{Additional resources}
\begin{itemize}
    \item Tutorial for building a java project with gradle: \href{https://docs.gradle.org/current/samples/sample_building_java_applications.html}{\faChain}.
    \item Tutorial for building a simple Java web application: \href{https://www.javahelps.com/2015/04/java-web-application-hello-world.html}{\faChain}
    \item Introduction to Spring Boot \href{https://www.tutorialspoint.com/spring_boot/spring_boot_introduction.htm}{\faChain}
    \item Introduction to Apache Wicket \href{https://wicket.apache.org/learn/examples/index.html}{\faChain}
    \item Tutorial for setting up a student Azure account: \href{https://www.learningjournal.guru/courses/modern-web-development/core-concepts/free-vm-in-azure/}{\faChain}
    \item Instructions for the installation of Oracle JDK 8 and configuration of the system environment variables on Linux: \href{https://www.javahelps.com/2015/03/install-oracle-jdk-in-ubuntu.html}{\faChain}
    \item Calling Java in Julia \href{https://juliainterop.github.io/JavaCall.jl/}{\faChain}, Python \href{https://jpype.readthedocs.io/en/latest/}{\faChain} and R \href{https://cran.r-project.org/web/packages/rJava/index.html}{\faChain}
\end{itemize}

\textbf{Installation}

If the correct version of Java is already installed and configured correctly, the user can skip step 1 below.
\begin{enumerate}
    \item CoinTossX is currently only compatible with version 8 of Java. Therefore, the user should install version 8 of either the Oracle (\href{https://www.oracle.com/za/java/technologies/javase/javase-jdk8-downloads.html}{\faChain}) or Open JDK (Java Development Kit)\footnote{Note that the user should install the JDK (which includes the JRE), not only the JRE.}. For simplicity, it is assumed that Windows and OSX users will install the Oracle JDK while Linux users will install the Open JDK.
    \begin{itemize}
        \item \textbf{Windows} --- After installing Oracle JDK 8 using the link above, if not done so automatically by the java install wizard, ensure that \lstinline[columns=fixed, language=custom]{JAVA_HOME} is set and that the java executable is set in the path environment variable. To set/add java to the system's \lstinline[columns=fixed, language=custom]{JAVA_HOME} and \lstinline[columns=fixed, language=custom]{path} environment variables, go to \lstinline[columns=fixed, language=custom]{Settings > Advanced System Settings > Environment Variables > System Variables}. Then add the location of the java installation to a new variable called \lstinline[columns=fixed, language=custom]{JAVA_HOME} which points to the relevant java distribution (e.g. \lstinline[columns=fixed, language=custom]{C:\Program Files\Java\jdk1.8.0_271}). Thereafter append a new pointer in the \lstinline[columns=fixed, language=custom]{path} environment variable by adding \lstinline[columns=fixed, language=custom]{\%JAVA_HOME\%\bin}.
        \item \textbf{Linux} --- Installation of Open JDK 8 is done by executing the commands below:
        \begin{lstlisting}[language = custom, numbers=none]
        sudo apt-get update
        sudo apt-get install openjdk-8-jdk
        \end{lstlisting}
        After this, if the correct version of Java is still not being used, the user can switch to the correct version using
        \begin{lstlisting}[language = custom, numbers=none]
        sudo update-alternatives --set java /usr/lib/jvm/jdk1.8.0_[version]/bin/java
        \end{lstlisting}
        \item \textbf{OSX} --- To set \lstinline[columns=fixed, language=custom]{JAVA_HOME} run
        \begin{lstlisting}[language = custom, numbers=none]
        export JAVA_HOME=/Library/Java/JavaVirtualMachines/jdk1.8.0_[version].jdk/Contents/Home
        \end{lstlisting}
        \item Clone the CoinTossX repository (using either \lstinline[columns=fixed, language=custom]{git} in the terminal or Github Desktop). The user can clone the repository to any location of their choosing.
        \begin{lstlisting}[language = custom, numbers=none, literate={.git}{.git}4]
        cd [directory to clone into]
        git clone https://github.com/dharmeshsing/CoinTossX.git
        \end{lstlisting}
    \end{itemize}
\end{enumerate}
\end{minipage}

The next set of instructions provides guidelines for the deployment of CoinTossX. By this point the user will have cloned the repository to a location of their choosing for either local or remote deployment. Depending on the operating system, different deployment configurations would need to be employed - hence the multiple configuration and deployment files. The files with the ".properties" extensions define the required ports for each component as well as user specific path definitions. To match the location to which the repository was cloned, the user would have to configure the \lstinline[columns=fixed, language=custom]{local.properties} file (for Linux) or the \lstinline[columns=fixed, language=custom]{windows.properties} file (for Windows) to correspond with the user's directories. For remote deployment the user would have to configure the \lstinline[columns=fixed, language=custom]{remote.properties} with the corresponding paths on the remote server. The path variables which need to be configured are:
\begin{itemize}
    \item \lstinline[columns=fixed, language=custom]{MEDIA_DRIVER_DIR} pointing to the Aeron Media Driver. This folder will only be created after the application is started. For Linux users it is recommended that the default path (\lstinline[columns=fixed, language=custom]{/dev/shm}) is not deviated from in order to achieve optimal performance. Otherwise the path can be amended as follows:
    \begin{lstlisting}[language = custom, numbers=none]
    MEDIA_DRIVER_DIR=[...]/aeron
    \end{lstlisting}
    Note that if the user deviates from the default media driver path, they would have to make the same change in the \lstinline[columns=fixed, language=custom]{build.gradle} file.
    \item \lstinline[columns=fixed, language=custom]{SOFTWARE_PATH} pointing to the start-up folder that will be created upon deployment. This path can be set to any valid path on the user's machine and may be given any name. For example:
    \begin{lstlisting}[language = custom, numbers=none]
    SOFTWARE_PATH=[...]/[folder name]
    \end{lstlisting}
    \item \lstinline[columns=fixed, language=custom]{DATA_PATH} points to the data folder within the software path. For example:
    \begin{lstlisting}[language = custom, numbers=none]
    DATA_PATH=[SOFTWARE_PATH]/data
    \end{lstlisting}
\end{itemize}

Before the application can be started, we are required to change a few system settings to ensure that network performance and system memory are utilised correctly. Firstly, the receive and send UDP buffer sizes/limits need to be configured as follows.
\begin{lstlisting}[language = custom, numbers=none]
sudo sysctl net.core.rmem_max=2097152
sudo sysctl net.core.wmem_max=2097152
\end{lstlisting}
Secondly, this Java application requires large pages to be enabled. By default Linux does not enable any HugePages (portions of memory set aside to be solely used by the application). To determine the current HugePage usage, run \lstinline[columns=fixed, language=custom]{grep Huge /proc/meminfo}. Furthermore, the default HugePage size is 2MB (2048kB). So if the user wishes to enable approximately 20GB of HugePages (dedicated memory), this would require 10000 HugePages - so the command to run would be
\begin{lstlisting}[language = custom, numbers=none]
sudo sysctl vm.nr_hugepages=10000
\end{lstlisting}

Last is the running of the application. Users deploying the application to Microsoft Azure, CHPC, Wits Server or any other server may choose to do so locally or remotely. Remote deployment will require that the user specify the above paths to correspond with that of the remote server. The instructions below demonstrate both local and remote deployment. For users deploying remotely, one must first ensure that SSH is enabled on the server and that port 22 is open for the transfer of files. Additionally, the username, IP address and password fields in the \lstinline[columns=fixed, language=custom]{deploy_remote.gradle} file should match that of the server.

\begin{enumerate}
    \item Set the current directory to the location of CoinTossX:
    \begin{lstlisting}[language = custom, numbers=none]
    cd [path to]/CoinTossX
    \end{lstlisting}
    \item To build the project, use the provided gradle wrapper to run:
    \begin{lstlisting}[language = custom, numbers=none]
    ./gradlew -Penv=local build -x test # For local deployment
    ./gradlew -Penv=remote build -x test # For remote deployment
    ./gradlew.bat -Penv=windows build -x test # For windows deployment
    \end{lstlisting}
    \item To deploy the project to a start-up folder, execute one of the following. 
    \begin{lstlisting}[language = custom, numbers=none]
    ./gradlew -Penv=local clean installDist bootWar copyResourcesToInstallDir copyToDeploy deployLocal # For local deployment
    ./gradlew -Penv=remote clean installDist bootWar copyResourcesToInstallDir copyToDeploy deployRemote # For remote deployment
    ./gradlew.bat -Penv=windows clean installDist bootWar copyResourcesToInstallDir copyToDeploy deployLocal # For windows deployment
    \end{lstlisting}
    \item Finally, to run the program, execute the shell script:
    \begin{lstlisting}[language = custom, numbers=none, literate={.sh}{.sh}3]
    sh [SOFTWARE_PATH]/scripts/startAll.sh # For local start-up
    sh [SOFTWARE_PATH]/scripts/remote_startAll.sh # For remote start-up
    [SOFTWARE_PATH]/scripts/startAll.bat # For Windows start-up
    \end{lstlisting}
    \item Access the web app by typing \lstinline[columns=fixed, language=custom]{localhost:8080} (if deployed locally) or \lstinline[columns=fixed, language=custom]{[server IP]:8080} (if deployed remotely) into the URL search bar of a browser.
\end{enumerate}

\textbf{Usage}

\begin{minipage}{.58\textwidth}
To configure the trading sessions to be fired during the simulation refer to \lstinline[columns=fixed, language=custom]{tradingSessionsCron.properties} file found in the \lstinline[columns=fixed, language=custom]{data} directory. The usage/syntax of the cron expressions within that file are as follows. A \href{https://docs.oracle.com/cd/E12058_01/doc/doc.1014/e12030/cron_expressions.htm}{cron expression} is a string consisting of six or seven fields, separated by white space, that describe individual details of the schedule:
\begin{lstlisting}[language = custom, numbers=none]
[seconds] [minutes] [hours] [day of month] [month] [day of week] [year]
\end{lstlisting}
A few examples, provided in \lstinline[columns=fixed, language=custom]{tradingSessionsCron.properties}, are shown on the right.
\end{minipage}
\begin{minipage}{.4\textwidth}
\begin{lstlisting}[language = custom, numbers=none]
TRADING_SESSIONS=ContinuousTrading2

#StartOfTrading.name=ContinuousTrading
#StartOfTrading.cron=0 0 7 * * 1-7

#ContinuousTrading.name=ContinuousTrading
#ContinuousTrading.cron=0 0 9 * * 1-7

#IntraDayAuctionCall.name=IntraDayAuctionCall
#IntraDayAuctionCall.cron=0 0 17 * * 1-7

ContinuousTrading2.name=ContinuousTrading
ContinuousTrading2.cron=0 15 17 * * 1-7

#RandomAuction.name=VolatilityAuctionCall
#RandomAuction.cron=0 0/5 * * * 1-7
\end{lstlisting}
\end{minipage}

Each component of the system as well as other actions can be started independently via the shell scripts. The list of all the runnable shell scripts can be found in the \lstinline[columns=fixed, language=custom]{deploy/scripts} directory. Each shell script simply executes the Java byte code of a ``main'' method whose class is specified in the \lstinline[columns=fixed, language=custom]{build.gradle} file of the corresponding module. For example, the \lstinline[columns=fixed, language=custom]{startAll.sh} script starts each component consecutively (equivalent to clicking the ``Start'' button on the Hawkes simulation web page). Similarly, \lstinline[columns=fixed, language=custom]{stopAll.sh} stops all the components (equivalent to clicking the ``Shut Down'' button on the Hawkes simulation webpage). Furthermore, any other actions on the website can also be done from the command line. To start the Hawkes simulation for a single client and stock simply run \lstinline[columns=fixed, language=custom]{./startHawkesSimulation.sh 1 1}. This starts the simulation for client 1 and stock 1 by submitting the client ID and stock ID as the first and second argument, respectively.

The list of clients that can be activated can be found in the \lstinline[columns=fixed, language=custom]{data/ClientData.csv} file. Consider, for example, the first client shown below.
\begin{lstlisting}[language = custom, numbers=none]
CompID,Password,NGInputURL,NGInputStreamId,NGOutputURL,NGOutputStreamId,MDGInputURL,MDGInputStreamId,MDGOutputURL, MDGOutputStreamId,SecurityId
1,test111111,udp://localhost:5000,10,udp://localhost:5001,10,udp://localhost:5002,10,udp://localhost:5003,10,1
\end{lstlisting}
Each client is assigned input and output URLs for both the Native Gateway and Market Data Gateway. These URLs specify the IP addresses (in this case localhost - the user's local machine) to/from which messages will be sent/received as well as the ports on which these components are listening.



From the website the user can: view/add/edit/delete clients, view the simulation status of clients and stocks, start the Hawkes simulation, edit the Hawkes simulation parameters, view/extract Hawkes simulation data as well as view a snapshot of the limit order books of each stock. With regards to the output of results, after all clients have submitted an end of trading session message, the orders along with their submission times are written to file and stored in the \lstinline[columns=fixed, language=custom]{deploy/data} directory. At the end of each Hawkes simulation the HdrHistogram latency results are written to a text file in the same directory.

\begin{figure}[H]
    \centering
    \settowidth{\imagewidth}{\includegraphics[height=3cm]{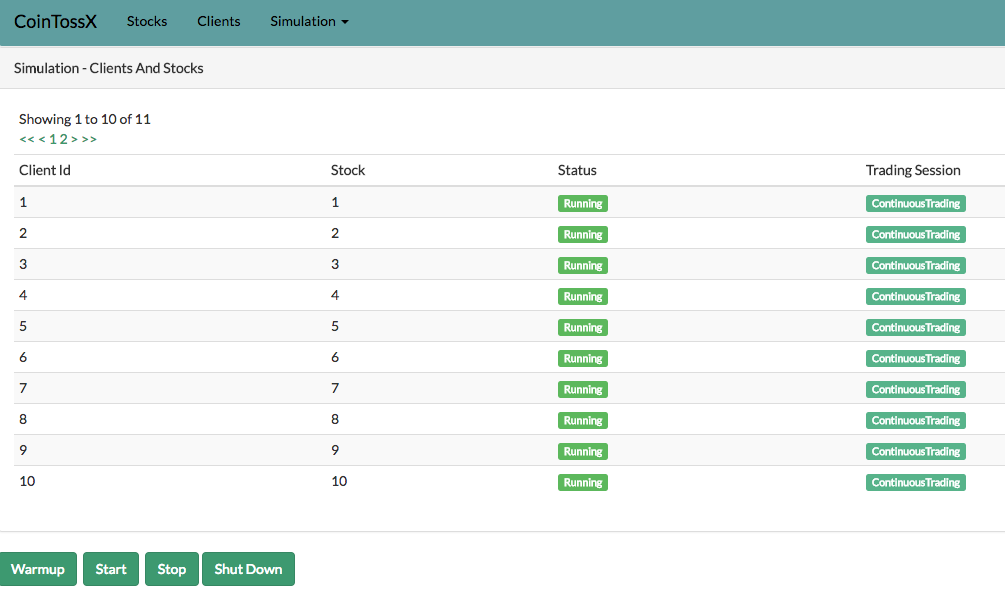}}
    \subfloat[Splash screen \label{figa:website 1}]{\makebox[\imagewidth][l]{\includegraphics[height=3cm]{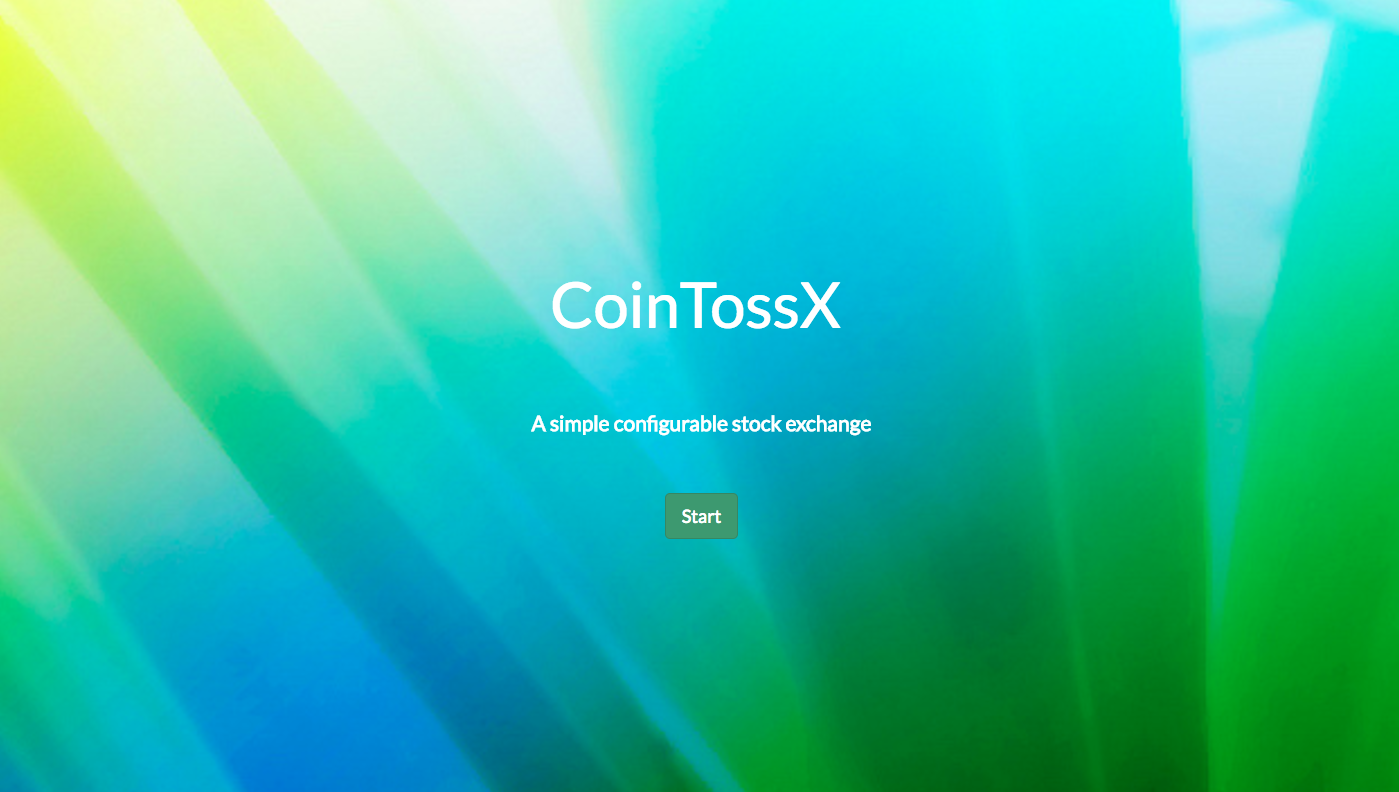}}} \quad
    \subfloat[All stocks and clients page snapshot \label{figb:website 2}]{\includegraphics[height=3cm]{Figures/CoinTossX_Website_2.png}} \quad
    \subfloat[Simulation page snapshot \label{figc:website 3}]{\makebox[\imagewidth][l]{\includegraphics[height=3cm]{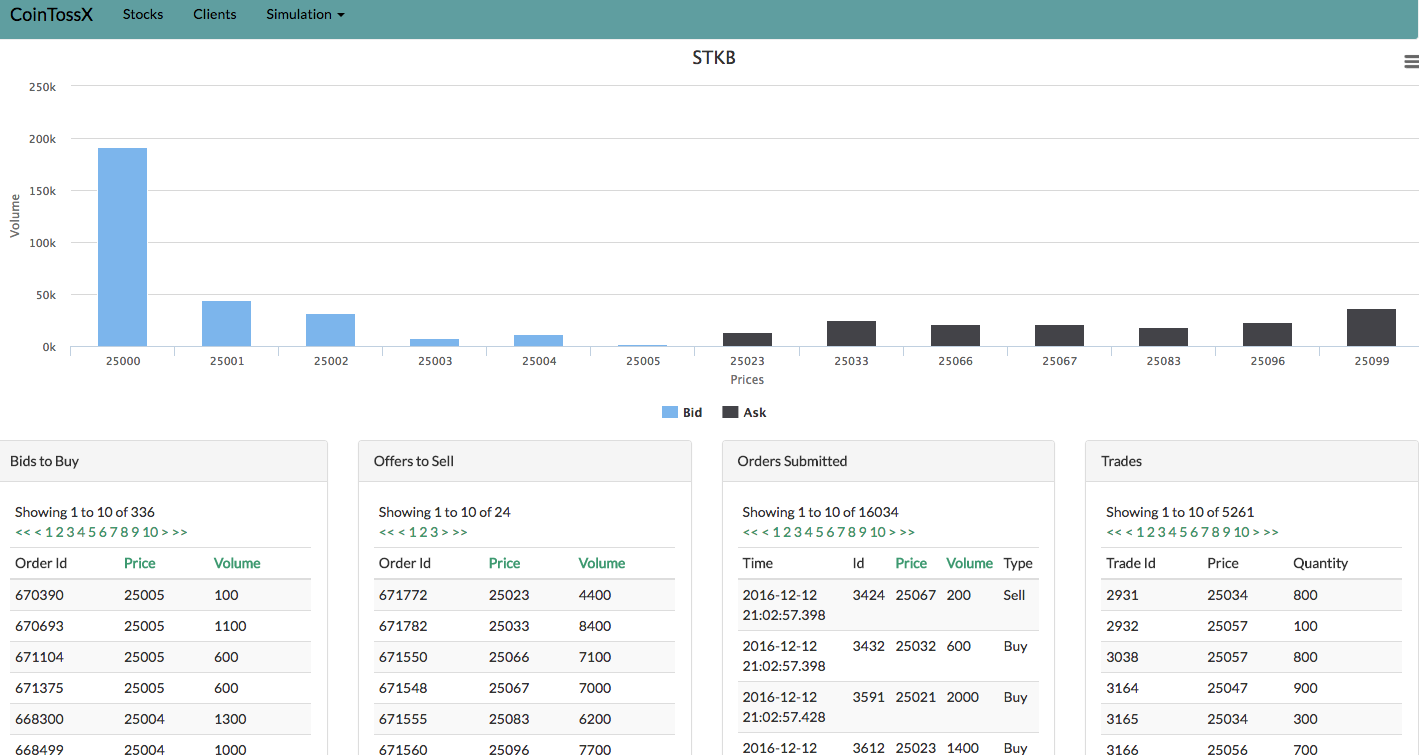}}} \\
    \subfloat[Hawkes configuration page snapshot \label{figd:website 4}]{\makebox[\imagewidth][l]{\includegraphics[height=3cm]{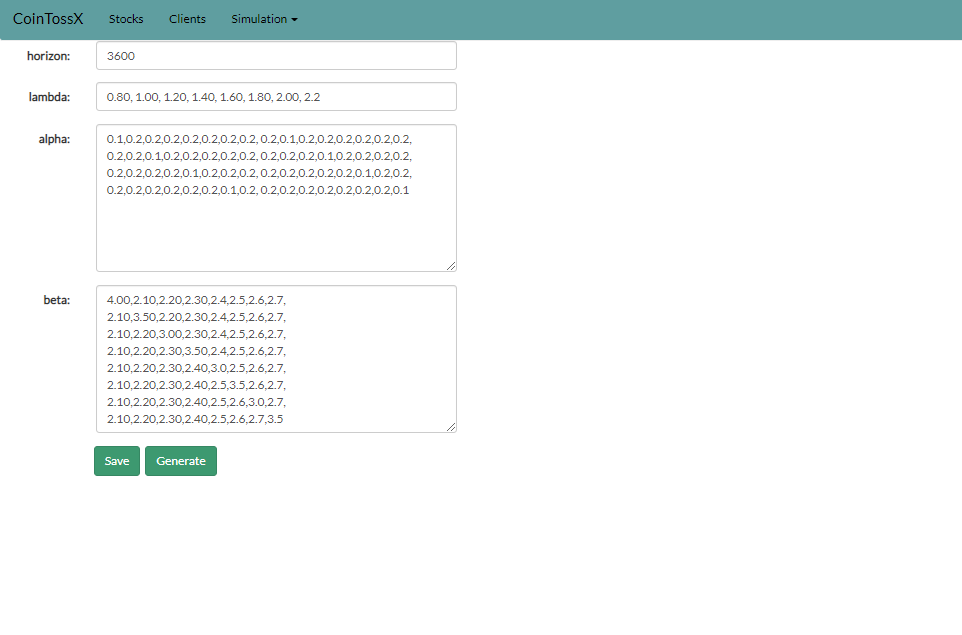}}} \quad
    \subfloat[Client page snapshot \label{fige:website 5}]{\makebox[\imagewidth][l]{\includegraphics[height=3cm]{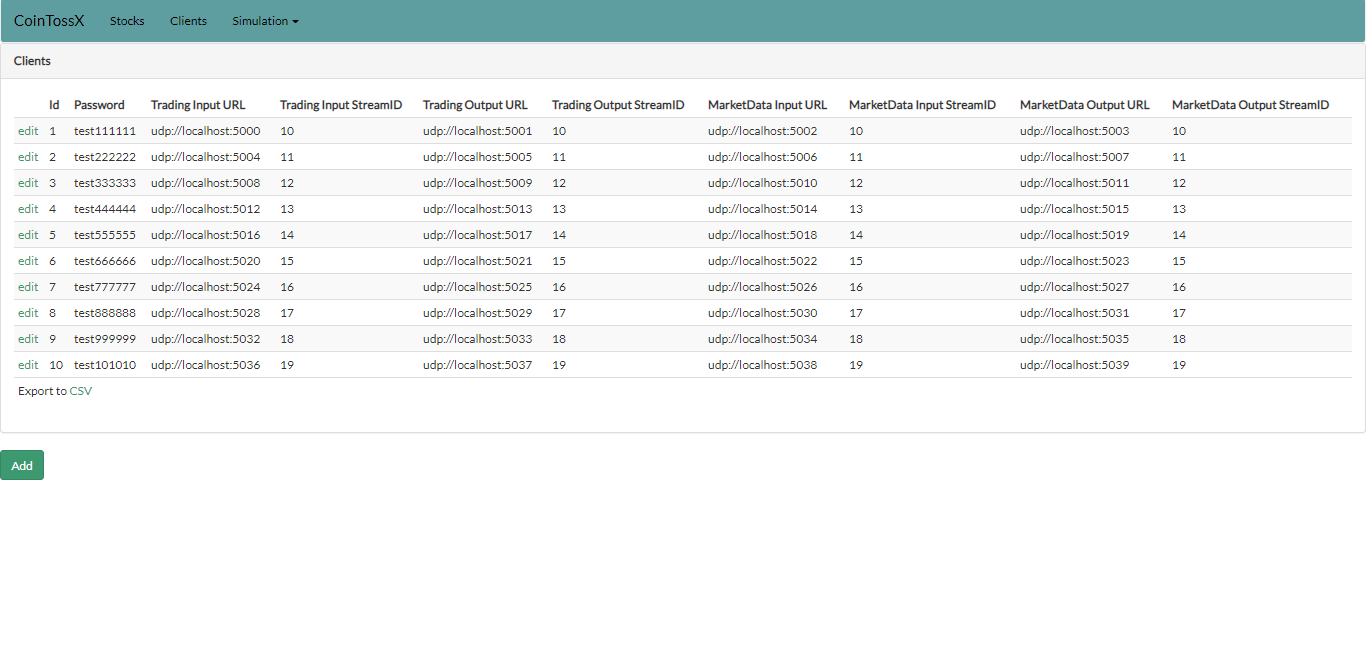}}}
    \caption{CoinTossX website \label{fig:website}}
\end{figure}

\section{Submitting orders \label{sec:Submitting orders}}
The subsections below demonstrate the submission of orders to the trading gateway (on a user's local machine) in three programming languages: Java, Julia and Python. Note that currently CoinTossX only allows for Java implementations. Nonetheless, most programming languages include packages/libraries which provide interfaces for calling Java. Although it may be considered sub-optimal, this solution allows us to bypass some problems with a Java-only implementation without the need for a software shim. Even though the message interface is port based, and hence it should not matter what language or interface is used, to submit orders the choices made here are convenient as they still allow the separation of the matching engine for the components that may generate orders. 

The below programming languages are only a few of those that provide these types of libraries. The code snippets below can be found in the CoinTossX project under the directory \lstinline[columns=fixed, language=custom]{ClientSimulator/src/main/java/example}. More specifically, all necessary static Java methods are defined in the \lstinline[columns=fixed, language=custom]{Utilities} class. Clients have the following functionality that will be demonstrated in the code snippets:
\begin{enumerate}
    \item Submit the order combinations shown in table \ref{table:session tif/order combinations}.
    \item Cancel limit orders.
    \item Market data updates
    \begin{enumerate}
        \item VWAP of buy/sell side of LOB.
        \item Current best bid/ask price/quantity.
        \item Active trading session.
        \item Whether or not the LOB has updated.
    \end{enumerate}
\end{enumerate}

\subsection{Java}
As opposed to the other implementations below, the Java implementation is shown in full without reference to the \lstinline[columns=fixed, language=custom]{Utilities} class to provide more detail for how the other implementations function.
\begin{lstlisting}[language = Java]
/*
Example:
- Java version: 1.8.0_271
- Authors: Ivan Jericevich, Dharmesh Sing, Tim Gebbie
- Structure:
    1. Dependencies
    2. Supplementary methods
    3. Implementation
        - Login and start session
        - Submit orders
        - Market data updates
        - End session and logout
*/
//---------------------------------------------------------------------------------------------------

//----- Import the necessary dependencies -----//
package example;

import client.*;
import sbe.msg.*;
import java.util.Properties;
import java.io.IOException;
import java.io.InputStream;
import java.util.Properties;
import java.time.LocalDateTime;
import com.carrotsearch.hppc.IntObjectMap;
//---------------------------------------------------------------------------------------------------

public class Example {
    private static final String PROPERTIES_FILE =  "simulation.properties"; // The name of the file specifying the simulation configuration
    private static Properties properties = new Properties();

    //----- Supplementary method for extracting the simulation settings -----//
    private static void loadProperties(Properties properties, String propertiesFile) throws IOException {
        try (InputStream inputStream = Example.class.getClassLoader().getResourceAsStream(propertiesFile)) {
            if (inputStream != null) {
                properties.load(inputStream);
            } else {
                throw new IOException("Unable to load properties file " + propertiesFile);
            }
        }
    }
    //---------------------------------------------------------------------------------------------------

    //----- Implementation -----//
    public static void main (String[] args) throws Exception {

        //----- Login and start session -----//
        int clientId = 1; int securityId = 1; // Define the client ID corresponding the client to be logged in as well as the security ID corresponding to the security in which they will trade
        // Load the simulation settings as well as all client data (ports, passwords and IDs)
        loadProperties(properties, PROPERTIES_FILE);
        String dataPath = properties.get("DATA_PATH").toString();
        IntObjectMap<ClientData> clientData = ClientData.loadClientDataData(dataPath);
        // Create and initialize client
        Client client = new Client(clientData.get(clientId), securityId);
        client.init(properties);
        // Start trading session by Logging in client to the gateways and connecting to ports
        client.sendStartMessage();
        System.out.println("Start at " + LocalDateTime.now());

        //----- Submit orders -----//
        // Arguments for "submitOrder": volume, price, side, order type, time in force, display quantity, min execution size, stop price
        client.submitOrder(1000, 99, "Buy", "Limit", "Day", 1000, 0, 0); // Buy limit order
        client.submitOrder(1000, 101, "Sell", "Limit", "Day", 1000, 0, 0); // Sell limit order
        client.submitOrder(1000, 0, "Buy", "Market", "Day", 1000, 0, 0); // Buy market order
        client.submitOrder(1000, 0, "Buy", "StopLimit", "Day", 1000, 0, 0); // Stop buy limit order
        client.submitOrder(1000, 0, "Buy", "Stop", "Day", 1000, 0, 0); // Stop buy market order
        client.cancelOrder("1", "Buy"); // Cancel limit order

        //----- Market data updates -----//
        client.calcVWAP("Buy"); // VWAP of buy side of LOB
        client.getBid(); // Best bid price
        client.getBidQuantity(); // Best bid volume
        client.getOffer(); // Best ask price
        client.getOfferQuantity(); // Best ask volume
        client.waitForMarketDataUpdate(); // Pauses the client until a new event occurs
        client.isAuction(); // Current trading session

        //----- End trading session by logging out client and closing connections -----//
        client.sendEndMessage();
        client.close();
        System.out.println("Complete at " + LocalDateTime.now());

        System.exit(0);
    }
    //------------------------------------------------------------------------------------------------------------------
}
\end{lstlisting}

\subsection{Julia}
The Julia-Java interface is provided in the \href{https://github.com/JuliaInterop/JavaCall.jl}{JavaCall.jl} package. This package works by specifying the paths to the compiled Java byte code (.jar files) as well as the path to the required dependencies. Thereafter the JVM is started and the utilities for logging a client in, starting the trading session, submitting an order and logging out are imported. With the execution of each command, the Java stacktrace is printed in the REPL as well.
\begin{lstlisting}[language = Julia]
#=
Example:
- Julia version: 1.5.3
- Java version: 1.8.0_271
- Authors: Ivan Jericevich, Dharmesh Sing, Tim Gebbie
- Structure:
    1. Preliminaries
    2. Login and start session
    3. Submit orders
    4. Market data updates
    5. End session and logout
=#
#---------------------------------------------------------------------------------------------------


#----- Preliminaries -----#
# Import the Java-Julia interface package
using JavaCall
cd(@__DIR__); clearconsole() # pwd()
# Add the path to java classes as well as the path to the ".jar" files containing the required java dependencies
JavaCall.addClassPath("/home/ivanjericevich/CoinTossX/ClientSimulator/build/classes/main") # JavaCall.getClassPath()
JavaCall.addClassPath("/home/ivanjericevich/CoinTossX/ClientSimulator/build/install/ClientSimulator/lib/*.jar")
# Initialize JVM
JavaCall.init()
# Import the class containing the reqired methods
utilities = @jimport example.Utilities # JavaCall.listmethods(utilities)
#---------------------------------------------------------------------------------------------------


#----- Login and start session -----#
# Define the client ID corresponding the client to be logged in as well as the security ID corresponding to the security in which they will trade
clientId = 1; securityId = 1
# Load the simulation settings as well as all client data (ports, passwords and IDs) and create/initialize client
client = jcall(utilities, "loadClientData", JavaObject{Symbol("client.Client")}, (jint, jint), clientId, securityId)
# Start trading session by Logging in client to the gateways and connecting to ports
jcall(client, "sendStartMessage", Nothing, ())
#---------------------------------------------------------------------------------------------------


#----- Submit orders -----#
# Arguments for "submitOrder": volume, price, side, order type, time in force, display quantity, min execution size, stop price
jcall(client, "submitOrder", Nothing, (jlong, jlong, JString, JString, JString, jlong, jlong, jlong), 1000, 99, "Buy", "Limit", "Day", 1000, 0, 0) # Buy limit order
jcall(client, "submitOrder", Nothing, (jlong, jlong, JString, JString, JString, jlong, jlong, jlong), 1000, 101, "Sell", "Limit", "Day", 1000, 0, 0) # Sell limit order
jcall(client, "submitOrder", Nothing, (jlong, jlong, JString, JString, JString, jlong, jlong, jlong), 1000, 0, "Buy", "Market", "Day", 1000, 0, 0) # Buy market order
jcall(client, "submitOrder", Nothing, (jlong, jlong, JString, JString, JString, jlong, jlong, jlong), 1000, 0, "Buy", "StopLimit", "Day", 1000, 0, 0) # Stop buy limit order
jcall(client, "submitOrder", Nothing, (jlong, jlong, JString, JString, JString, jlong, jlong, jlong), 1000, 0, "Buy", "Stop", "Day", 1000, 0, 0) # Stop buy market order
# Arguments for "cancelOrder": order id, side
jcall(client, "cancelOrder", Nothing, (jlong, jlong), "1", "Buy") # Cancel limit order
#---------------------------------------------------------------------------------------------------


#----- Market data updates -----#
jcall(client, "calcVWAP", jlong, (JString,), "Buy") # VWAP of buy side of LOB
jcall(client, "getBid", jlong, ()) # Best bid price
jcall(client, "getBidQuantity", jlong, ()) # Best bid volume
jcall(client, "getOffer", jlong, ()) # Best ask price
jcall(client, "getOfferQuantity", jlong, ()) # Best ask volume
jcall(client, "waitForMarketDataUpdate", Nothing, ()) # Pauses the client until a new event occurs
jcall(client, "isAuction", jboolean, ()) # Current trading session
#---------------------------------------------------------------------------------------------------


#----- End session and logout -----#
# End trading session by logging out client and closing connections
jcall(client, "sendEndMessage", Nothing, ()); jcall(client, "close", Nothing, ())
#---------------------------------------------------------------------------------------------------
\end{lstlisting}

\subsection{Python}
The Python Java interface is provided in the \href{https://github.com/jpype-project/jpype}{JPype} module. This code snippet follows a similar process as above but with different syntax.
\begin{lstlisting}[language = Python]
'''
Example:
- Python version: 3.8.5
- Java version: 1.8.0_271
- Authors: Ivan Jericevich, Dharmesh Sing, Tim Gebbie
- Structure:
    1. Preliminaries
    2. Login and start session
    3. Submit orders
    4. Market data updates
    5. End session and logout
'''
#---------------------------------------------------------------------------------------------------


#----- Preliminaries -----#
# Import the Java-Python interface module
import jpype as jp
# Initialize/start the JVM
jp.startJVM(jp.getDefaultJVMPath(), "-ea", classpath = "/home/ivanjericevich/CoinTossX/ClientSimulator/build/classes/main") # Start JVM
# Add the path to the ".jar" files containing the required java dependencies
jpype.addClassPath("/home/ivanjericevich/CoinTossX/ClientSimulator/build/install/ClientSimulator/lib/*.jar")
# Import the class containing the reqired methods
utilities = jp.JClass("example.Utilities")
#---------------------------------------------------------------------------------------------------


#----- Login and start session -----#
# Define the client ID corresponding the client to be logged in as well as the security ID corresponding to the security in which they will trade
clientId = 1
securityId = 1
# Load the simulation settings as well as all client data (ports, passwords and IDs) and create/initialize client
client = utilities.loadClientData(clientId, securityId)
# Start trading session by Logging in client to the gateways and connecting to ports
client.sendStartMessage()
#---------------------------------------------------------------------------------------------------


#----- Submit orders -----#
# Arguments for "submitOrder": volume, price, side, order type, time in force, display quantity, min execution size, stop price
client.submitOrder(1000, 99, "Buy", "Limit", "Day", 1000, 0, 0) # Buy limit order
client.submitOrder(1000, 101, "Sell", "Limit", "Day", 1000, 0, 0) # Sell limit order
client.submitOrder(1000, 0, "Buy", "Market", "Day", 1000, 0, 0) # Buy market order
client.submitOrder(1000, 0, "Buy", "StopLimit", "Day", 1000, 0, 0) # Stop buy limit order
client.submitOrder(1000, 0, "Buy", "Stop", "Day", 1000, 0, 0) # Stop buy market order
# Arguments for "cancelOrder": order id, side
client.cancelOrder("1", "Buy") # Cancel limit order
#---------------------------------------------------------------------------------------------------


#----- Market data updates -----#
client.calcVWAP("Buy") # VWAP of buy side of LOB
client.getBid() # Best bid price
client.getBidQuantity() # Best bid volume
client.getOffer() # Best ask price
client.getOfferQuantity() # Best ask volume
client.waitForMarketDataUpdate() # Pauses the client until a new event occurs
client.isAuction() # Current trading session
#---------------------------------------------------------------------------------------------------


#----- End session and logout -----#
# End trading session by logging out client and closing connections
client.sendEndMessage()
client.close()
# Stop JVM
jp.shutdownJVM()
#---------------------------------------------------------------------------------------------------
\end{lstlisting}

\end{document}